\documentclass[10pt]{iopart}
\usepackage{graphicx}
\usepackage{iopams}
\usepackage{amsmath,amsthm,amssymb,amsfonts}
\usepackage[normalem]{ulem}
\usepackage{color,soul}
\usepackage[margin=3cm]{geometry} 
\newcommand{\stkout}[1]{\ifmmode\text{\sout{\ensuremath{#1}}}\else\sout{#1}\fi}

\usepackage{float}
\def\hlMode{0} % 1 = highlight on; 0 = highlight off
\newcommand{\hlcolor}[2]{\sethlcolor{#1}\hl{#2}}

\newcommand{\hb}[1]{\hlcolor{cyan}{#1}}

\if\hlMode0
    \renewcommand{\hlcolor}[2]{#2}
    
\fi

\begin{document}
%\title[Short title]{Full title}
%\title[Resilient Stellarator Divertor Characteristics in HSX]{Resilient Stellarator Divertor Characteristics in the Helically Symmetric eXperiment}
\title{Resilient Stellarator Divertor Characteristics in the Helically Symmetric eXperiment}

\author{K.A. Garcia$^1$, A. Bader$^2$, D. Boeyaert$^1$, A.H. Boozer$^3$, H. Frerichs$^1$, M.J. Gerard$^1$, A. Punjabi$^4$, \& O. Schmitz$^1$}

\address{$^1$University of Wisconsin-Madison}
\address{$^2$Type One Energy Inc}
\address{$^3$Columbia University}
\address{$^4$Hampton University}
\ead{kgarcia26@wisc.edu}

\begin{abstract} Resilient divertor features connected to open chaotic edge structures in the Helically Symmetric Experiment (HSX) are investigated. For the first time, an expanded vessel wall was considered that would give space for implementation of a physical divertor target structure. The analysis was done for four different magnetic configurations with very different chaotic plasma edges. A resilient plasma wall interaction pattern was identified across all configurations. This manifests as qualitatively very similar footprint behavior across the different plasma equilibria. Overall, the resilient field lines of interest with high connection length $L_C$ lie within a helical band along the wall for all configurations. This resiliency can be used to identify the best location of a divertor. The details of the magnetic footprint's resilient helical band is subject to specific field line structures which are linked to the penetration depth of field lines into the plasma and directly influence the heat and particle flux patterns. The differences arising from these details are characterized by introducing a new metric, the minimum radial connection min$(\delta_N)$ of a field line from the last closed flux surface. The relationship, namely the deviation from a scaling law, between min$(\delta_N)$ and $L_C$ of the field lines in the plasma edge field line behavior suggests that the field lines are associated with structures such as resonant islands, cantori, and turnstiles. This helps determine the relevant magnetic flux channels based on the radial location of these chaotic edge structures and the divertor target footprint. These details will need to be taken into account for resilient divertor design. 

\end{abstract}
\submitto{\PPCF}
\maketitle
%\ioptwocol

\section{Introduction}
\label{section:intro}

The non-resonant divertor (NRD) is a divertor concept gaining traction as a viable candidate for stellarator fusion reactors. Current stellarator divertors include the island divertor concept which has been explored in W7-AS and W7-X \cite{grigull_first_2001, renner_divertor_2002, wolf_performance_2019} and the helical divertor in LHD \cite{ohyabu_large_1994}. The helical troughs found in stellarator edges \cite{strumberger_magnetic_1992} are exploited in the NRD concept where a resilient field line intersection pattern on plasma facing components (PFCs) is observed across varying plasma equilibria, such as in HSX \cite{bader_hsx_2017} and in CTH \cite{bader_minimum_2018,garcia_exploration_2023}. This wall intersection pattern is also known as the strike line or point pattern of the field lines and serves as a proxy for heat and particle flux. Along with this characteristic resilient pattern, NRDs have been computationally shown to make use of chaotic structures, such as turnstiles along with homoclinic and heteroclinic tangles, in the plasma edge. These structures guide particle and heat flux to the PFCs \cite{punjabi_magnetic_2022,boozer_magnetic_2023,garcia_exploration_2023}. Tangles arising from Hamiltonian systems have been found experimentally in Tore Supra \cite{nguyen_connexion_1990,nguyen_interaction_1997}, TEXTOR \cite{jakubowski_modelling_2004,finken_structure_2005,jakubowski_influence_2007,schmitz_plasma_2012}, and DIII-D \cite{evans_experimental_2005}. Experimental measurements in LHD \cite{drewelow_comparison_2013}, CTH  \cite{allen_aps_nodate}, and Heliotron J \cite{cai_impact_2024} also suggest the presence and influence of similar chaotic plasma edge structures. In all cases, they influence the details of the deposition on PFCs.

This work expands on previous NRD research specifically for HSX as done in \cite{bader_hsx_2017} by using the field line tracer FLARE \cite{frerichs_flare_2024}. To explore the chaotic structures present in the HSX plasma edge, a lofted wall \hb{\mbox{\cite{schmitt_vacuum_2025}}} is used for simulations rather than the current physical vessel wall of HSX. This lofted wall is the largest extent the current vessel wall can be expanded such that the wall remains within the physical coils. \hb{The advantage of using an expanded wall further away from the plasma core is that this increases the width of the plasma edge region where various topological structures, such as islands and flux tubes, across different equilibria can be explored further computationally.} The work presented in this manuscript adds to the fundamental understanding of how chaotic structures arising in plasma edge topology influence the details of the NRD characteristic resilient deposition pattern. 

HSX is the first and only stellarator experiment optimized for quasi-helical symmetry (QHS) \cite{anderson_helically_1995}. Investigating NRD properties in a QHS device is a key step for designing resilient divertors for future reactor-scale quasi-symmetric stellarators \cite{nuhrenberg_quasi-helically_1988,landreman_magnetic_2022}. Reactor-scale quasi-symmetric configurations will have self-generated bootstrap currents \cite{redl_new_2021,landreman_optimization_2022} which necessitate the drive for a resilient divertor structure. Even in configurations with low-bootstrap current such as quasi-isodynamic configurations, a functional non-resonant divertor may be able to reduce the requirements on the plasma edge, expanding the acceptable configuration space. The NRD has been shown to exhibit resilient properties in particular for HSX \cite{bader_hsx_2017}. Hence, this work aims to expand on this previous research with the addition of the aforementioned lofted wall to enable further plasma wall interaction (PWI) studies for NRDs. 

The layout of the paper is as follows: Section \ref{section:hsxedge} describes the varying chaotic edge topology across 4 magnetic configurations considered for the analysis in this paper. Special attention is dedicated to the PWI of the chaotic edge field line behavior in vicinity of the lofted wall as we explore the characteristic resilient strike line pattern of NRDs in HSX. A new metric, the radial connection min$(\delta_N)$, is introduced to aid in this analysis. Section \ref{section:differences} then uses min$(\delta_N)$ to suggest the field lines behavior associated with islands and other chaotic edge structures which influence the PWI across the different plasma equilibria. Finally, section \ref{section:disc} discusses the impact of the different PWI for understanding NRDs and summarizes this work. 

\section{HSX Edge Structure}
\label{section:hsxedge}

In this section we will look at two methods for characterizing the edge of HSX with the lofted wall. We first introduce the 4 different plasma equilibria in this study followed by tracing the field lines in the vicinity of the wall to examine the edge PWI. It is often the case that looking solely at strike point calculations, as was done in \cite{bader_hsx_2017}, can appear to show identical behavior across several configurations. However, this approach may mask some edge features. The metrics calculated in this section will together show differences in the PWI in section \ref{section:differences} which will be connected to different plasma edge structures.

\subsection{Chaotic Layer Variation Across 4 Magnetic Equilibria}
\label{subsection:chaotic_layer_variation}

To analyze the plasma edge topology in HSX across various magnetic equilibria, we perform field line following with the FLARE code \cite{frerichs_flare_2024}. FLARE requires a magnetic model and defined boundary. For the magnetic model, the different plasma equilibria were produced by the 3D magnetohydrodynamic (MHD) code VMEC \cite{hirshman_momcon_1986} in free boundary mode. 
The lofted vessel wall \hb{\mbox{\cite{schmitt_vacuum_2025}}} is implemented as an extended boundary to analyze the topological structures present in the plasma edge of the configurations considered in this work. 

\begin{figure}[H]
    \centering
    \includegraphics[scale=0.2]{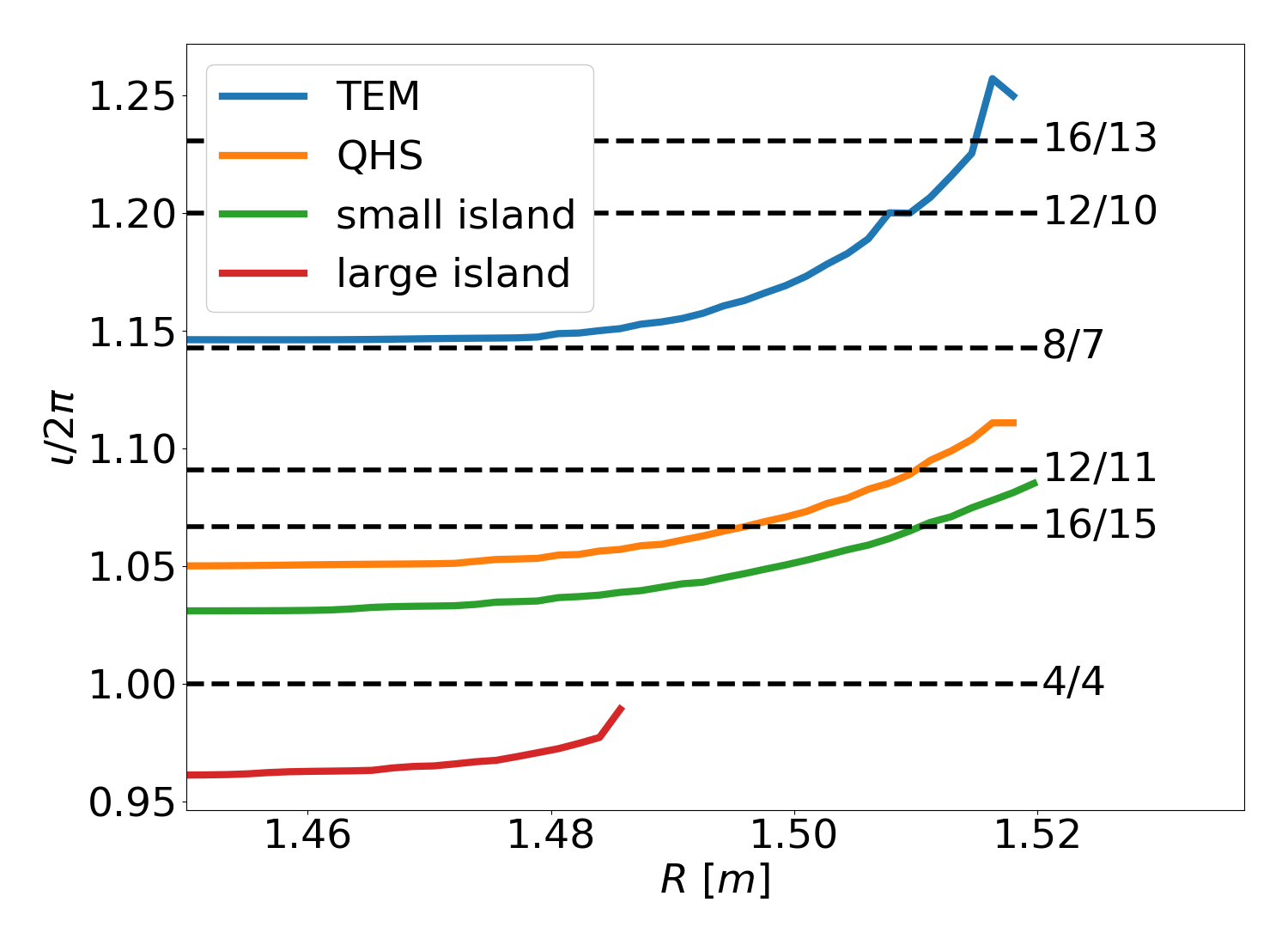}
    \caption{Rotational transform profile for the TEM (blue), QHS (orange), small island (green), and large island (red) magnetic configurations. The dashed black lines indicate different $m/n$ rational surfaces.}
    \label{fig:iota}
\end{figure}

\begin{figure}[H]
    \centering
    \includegraphics[scale=0.255]{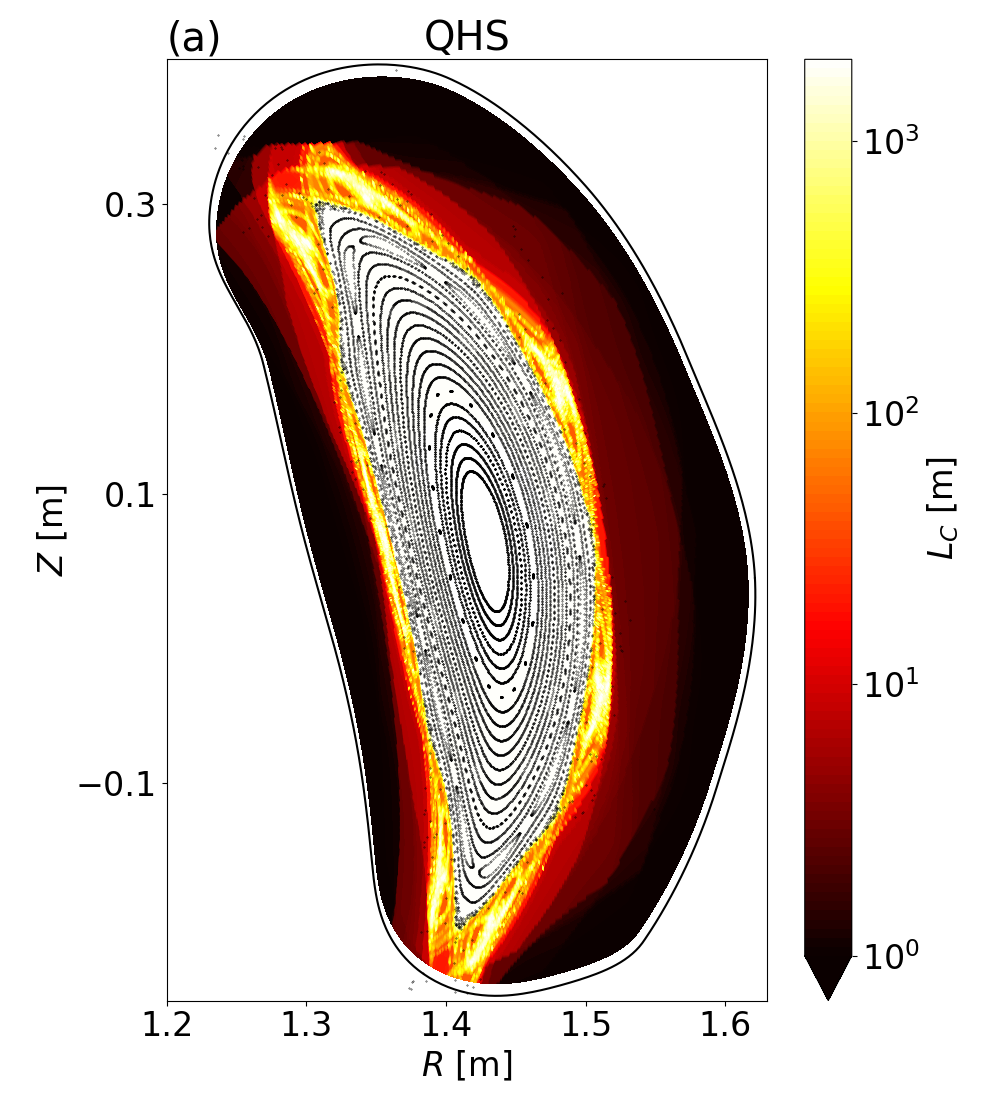}
    \includegraphics[scale=0.255]{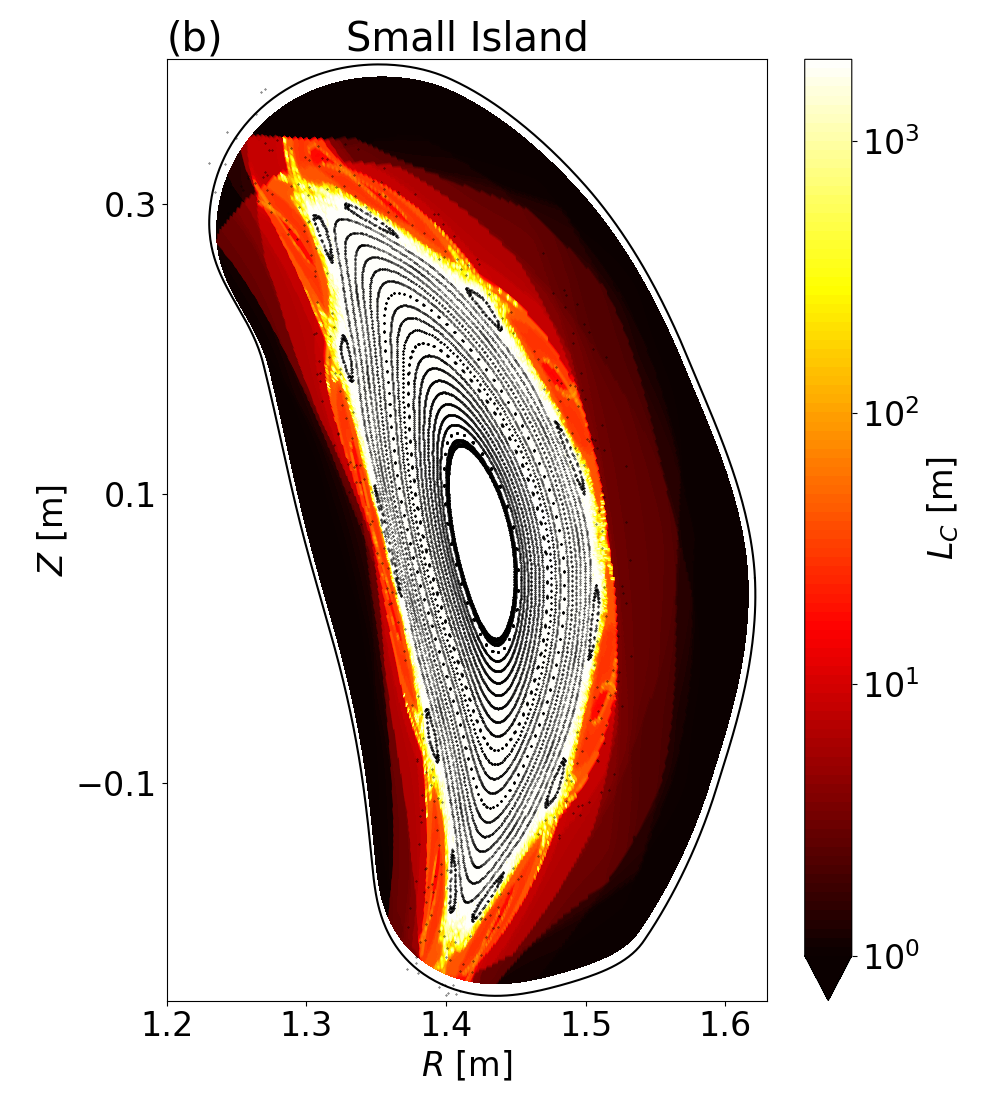} 
    \includegraphics[scale=0.255]{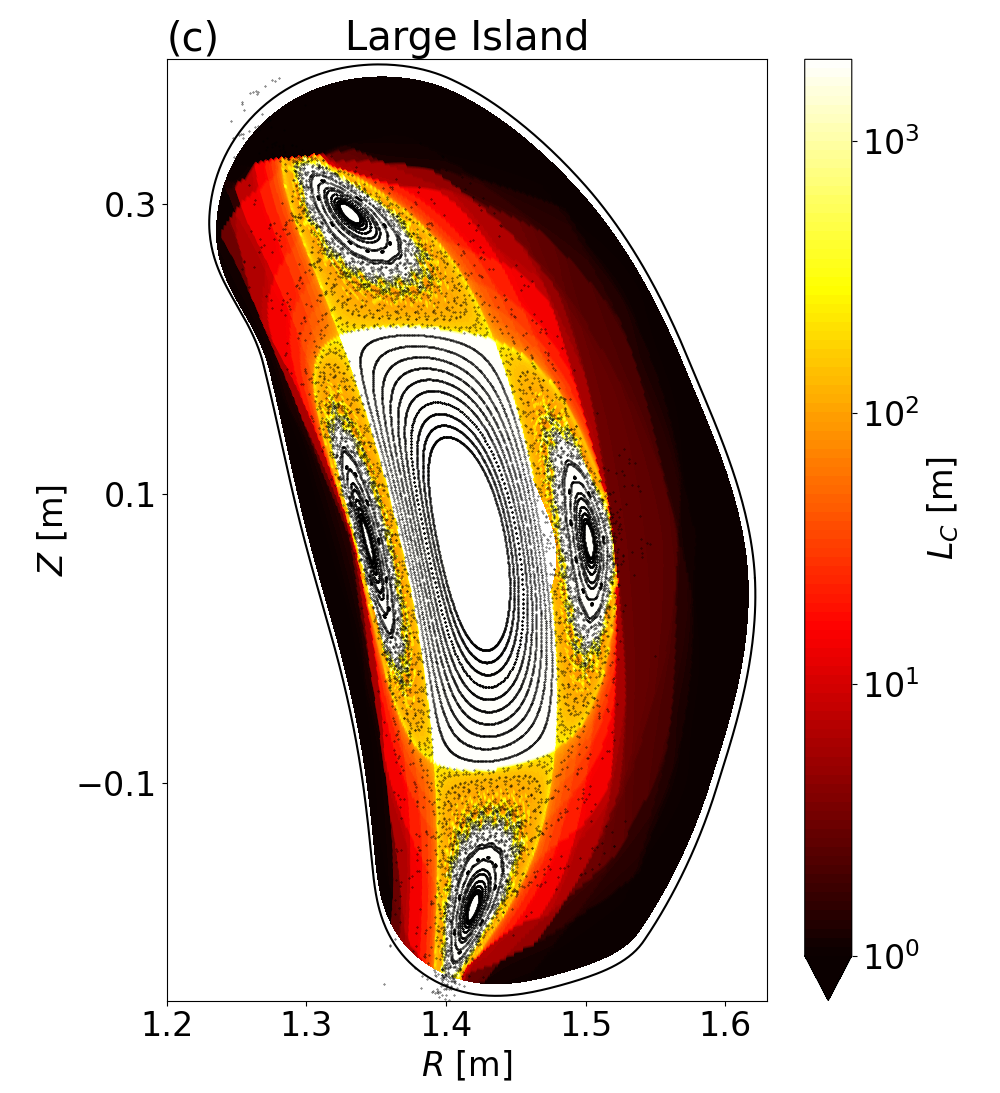}
    \includegraphics[scale=0.255]{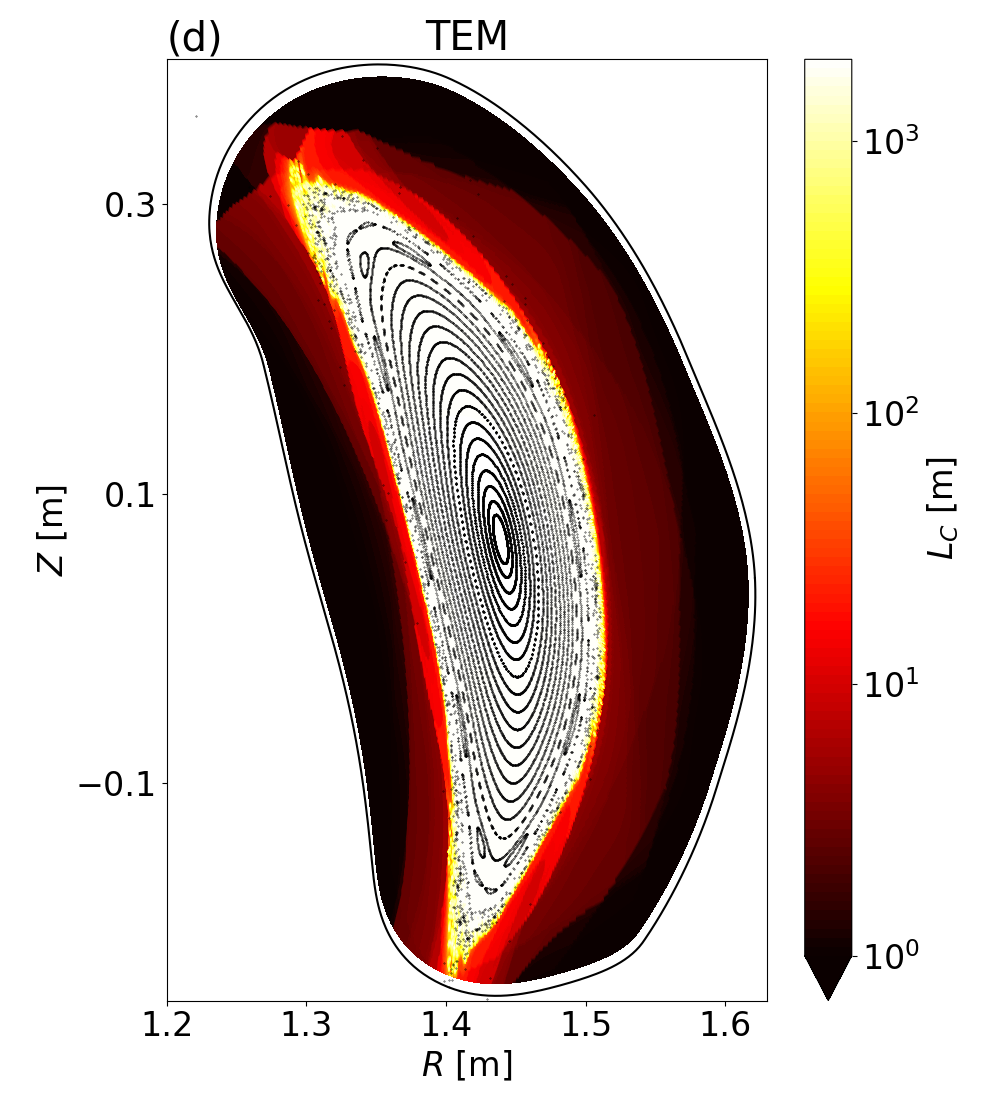}
    \caption{Connection length $L_C$ contour map and Poincar\`e plots at $\phi=5^{\circ}$ for 4 different equilibria: (a) QHS, (b) small island, (c) large island, and (d) TEM. }
    \label{fig:poincare5_Lc}
\end{figure}

\begin{figure}[H]
    \centering
    \includegraphics[scale=0.255]{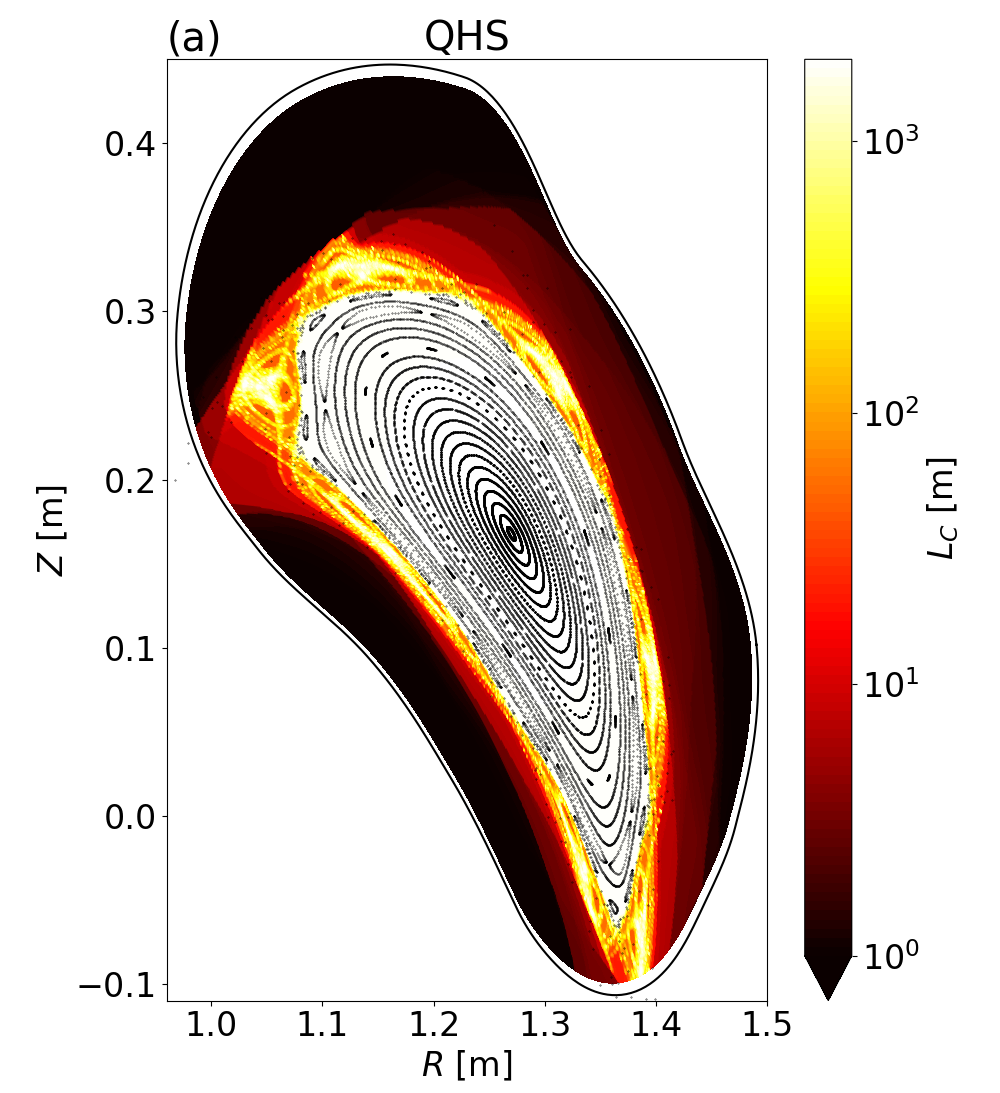}
    \includegraphics[scale=0.255]{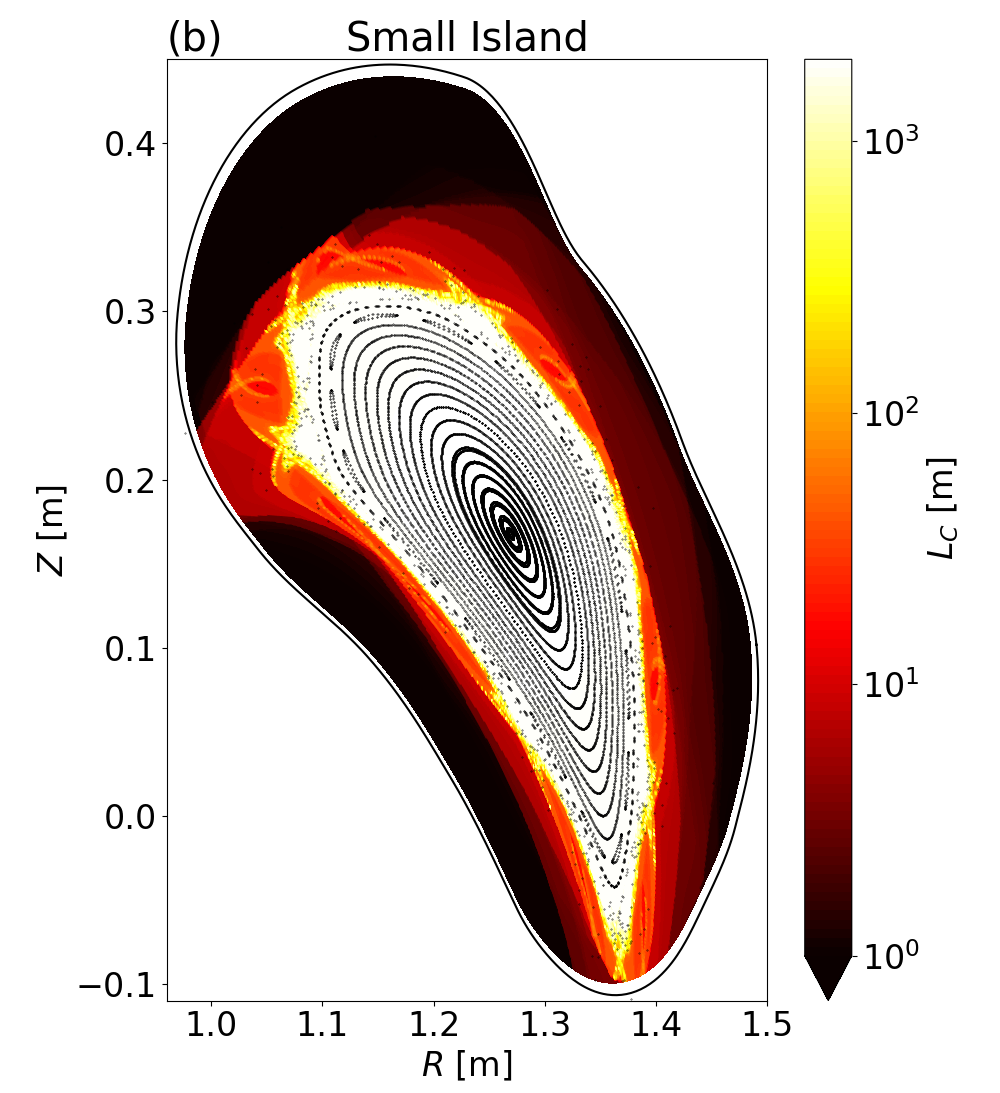} 
    \includegraphics[scale=0.255]{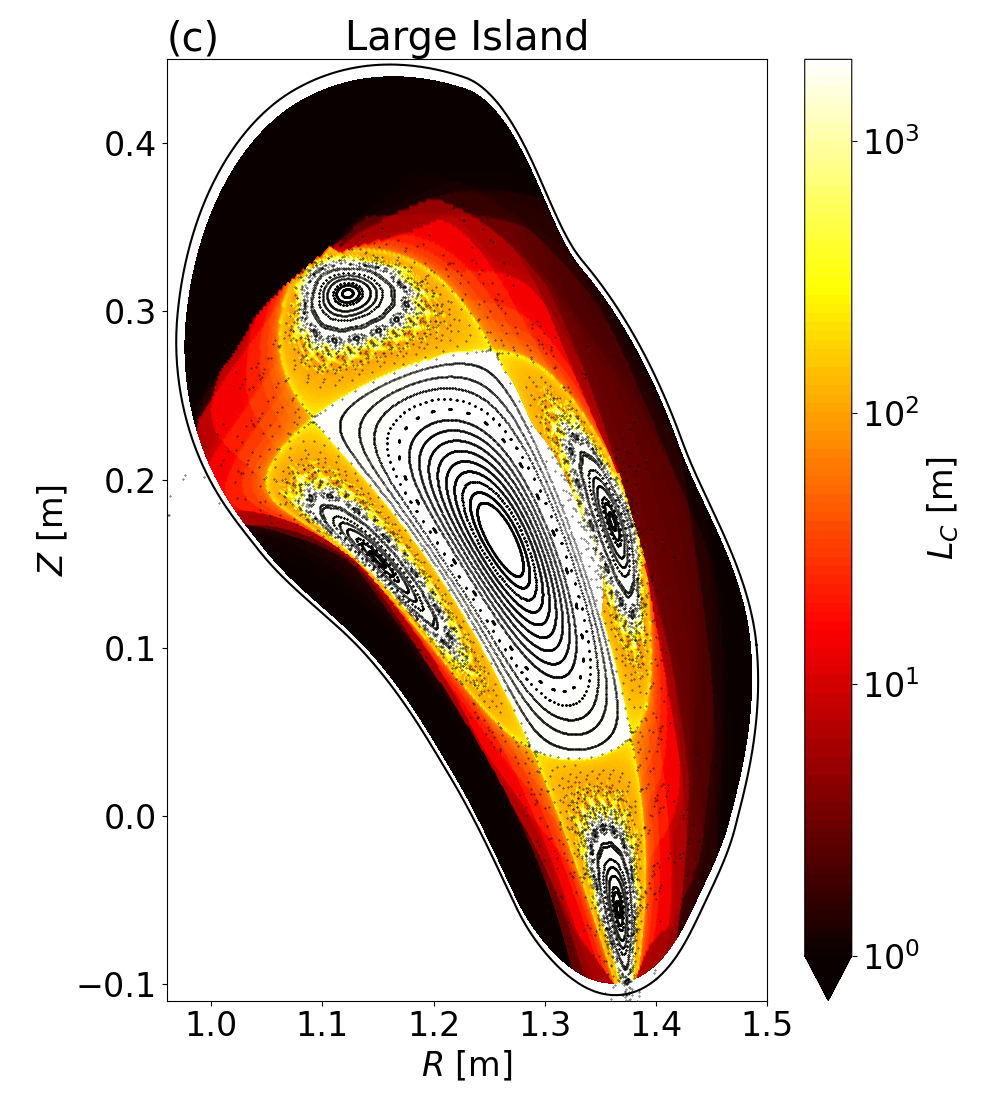}
    \includegraphics[scale=0.255]{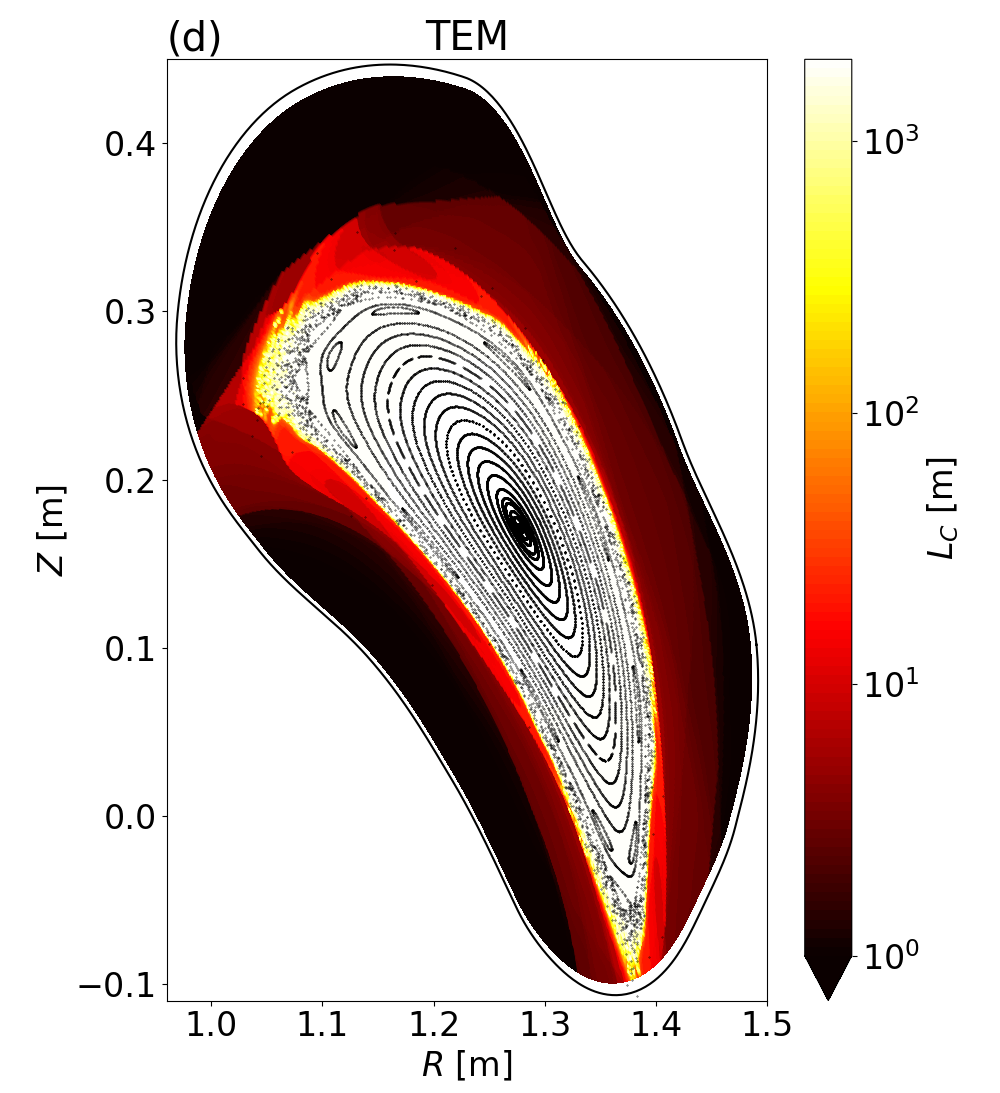}
    \caption{Connection length $L_C$ contour map and Poincar\`e plots at $\phi=18^{\circ}$ for 4 different equilibria: (a) QHS, (b) small island, (c) large island, and (d) TEM. }
    \label{fig:poincare18_Lc}
\end{figure}

HSX is a 4-field period stellarator with a minor radius of $a = 0.12 \ $m, a major radius of $R = 1.2 \ $m, and an aspect ratio of $R/a=10$. Figure \ref{fig:iota} provides a rotational transform profile $\iota$ inside the last closed flux surface (LCFS) as a function of $R$ of the 4 magnetic configurations considered in this paper: the standard quasi-helically \hb{symmetric} configuration (QHS), the small island configuration, the large island configuration, and the reduced trapped electron mode configuration (TEM) developed in \cite{gerard_optimizing_2023}. \hb{The first 3 configurations are equilibria which are similar to the ones studied in reference\mbox{\cite{bader_hsx_2017}} which enables comparison with previous results but with a different wall location. The TEM case is a recently developed configuration due to the growing interest in turbulence optimized configurations \mbox{\cite{gerard_optimizing_2023}} where the edge field line behavior has yet to be explored.} The dashed lines in black indicate the different $m/n$ rational surfaces, where $m$ and $n$ are integers. The large island configuration in red has the smallest core plasma volume because $\iota$ approaches the $4/4$ rational surface in the edge as seen in figure \ref{fig:iota}. This case is characterized by the presence of the $4/4$ island chain which can be seen in figures \ref{fig:poincare5_Lc} (c) and \ref{fig:poincare18_Lc} (c). \hb{It will be shown that the presence of the $4/4$ island chain and its PWI with the lofted wall makes this case the most ``island-divertor-like" configuration in contrast to the others.} Next, the small island and QHS cases in green and orange respectively both have $\iota$ which cross the $16/15$ rational surfaces and approach higher order rational surfaces, such as the $8/7$ island chain. This will be discussed more in figures \ref{fig:poincare5_Lc} (a)-(b) and \ref{fig:poincare18_Lc} (a)-(b) where the \hb{location of the X and O-points of the } $8/7$ islands are plotted \hb{and influence the PWI to be studied later}. Finally, the TEM case in blue has the highest $\iota$ in contrast to the other 3 configurations along with an increased plasma volume. Figure \ref{fig:iota} shows the higher order rational surfaces that this profile crosses. The details about this magnetic configuration can be found in reference \cite{gerard_optimizing_2023}. All cases of these configurations show that $\iota$ tends to increase especially toward the higher plasma edge values of $R$.

Figures \ref{fig:poincare5_Lc} and \ref{fig:poincare18_Lc} show contour maps of the connection length ($L_C$) in logarithmic scale for $\phi=5^{\circ}$ and $\phi=18^{\circ}$ planes, respectively, of the 4 different magnetic configurations where $\phi$ and $\theta$ are the toroidal and poloidal angles respectively. These two toroidal angles were chosen to highlight some specific features of the HSX edge which will be emphasized in section \ref{section:differences}. A Poincar\`e map is superimposed in black in each case along with the extended vessel wall. The connection length is the sum of the distances in the forward and backward directions along a field line until it strikes the lofted vessel wall. This computation was performed for a maximum $L_C$ of $1 \ $km in each direction, therefore, confined field lines that do not intersect the wall have maximum $L_C = 2 \ $km. This is shown as white in the contour plots. Field lines radially outward of the core have $L_C < 2 \ $km, and it is in this edge region of the plasma where the details of the $L_C$ give rise to differences in magnetic configuration topology. 

Starting with the large island magnetic configuration shown in figures \ref{fig:iota} (red), \ref{fig:poincare5_Lc} (c), and \ref{fig:poincare18_Lc} (c), the $4/4$ island chain dominates the plasma edge and the core plasma domain with closed flux surfaces is reduced. The Poincar\'e map superimposed on these figures shows that the islands are embedded in a region of chaotic field lines outside the LCFS. Around the islands in these figures, radially short collimated flux tubes of $L_C \leq 1 \ $km (yellow on the colorbar) overlap one another around the islands and create a chain of small secondary X-points. These are most visible in figures \ref{fig:poincare5_Lc_DeltaN} (f) and \ref{fig:poincare18_Lc_DeltaN} (f) to be discussed later. Each poloidal cross section in the two figures shows that the lofted wall intercepts islands in the edge making this configuration ``island divertor-like".  Although unlike a true island divertor like W7-X, the intersection between the island and the wall is limited to the very outer edge of the island. Nevertheless, island divertor-like features are discernable while exhibiting resilient behavior with the lofted wall. This will be shown later when discussing the PWI. 

Next, the small island configuration is discussed. The rotational transform is shown in green in figure \ref{fig:iota}, where the value approaches the $12/11$ and $8/7$ island chains at the edge. The $12/11$ island chain is visible in the Poincar\'e plot. The $8/7$ islands are also present in the edge. These islands are flattened and appear as red regions of medium connection length outside the confined plasma and intersect the wall as shown in figures \ref{fig:poincare5_Lc} (b) and \ref{fig:poincare18_Lc} (b). Small flux tubes of long $L_C$ (yellow) are most visible and present near the X-points and resemble divertor legs. Figures \ref{fig:poincare5_Lc} (b) and \ref{fig:poincare18_Lc} (b) show that the X-point nearest $R \sim 1.4 \ $m in the region of highest curvature has flux tubes (divertor legs) which are intercepted by the wall within a short radial distance. This behavior can be contrasted with the QHS magnetic configuration shown in \ref{fig:poincare5_Lc} (a), and \ref{fig:poincare18_Lc} (a). The QHS case plasma edge also features the $8/7$ island chain, as seen in figures \ref{fig:iota} (blue). Complicated internal structure of the island is visible with several regions of long, but not infinite, connection length. These flux tubes reach the vessel wall in the regions of high curvature corresponding to $R \sim 1.4 \ $m and minimum $Z$ in figures \ref{fig:poincare5_Lc} (a) and \ref{fig:poincare18_Lc} (a). This is similarly observed in the TEM configuration case in figures \ref{fig:iota} (orange), \ref{fig:poincare5_Lc} (d), and \ref{fig:poincare5_Lc} (d). However, due to the nature of the topology of this configuration, where no edge islands are visible of any size due to the island overlap of the $12/11$ and $16/13$ resonances, these flux tubes do not have a pronounced coherent shape like in the QHS and small island cases. 

\subsection{Magnetic Footprint on Lofted Wall}
\label{subsection:magfoot}

\begin{figure}[H]
    \centering
    \includegraphics[scale=0.182]{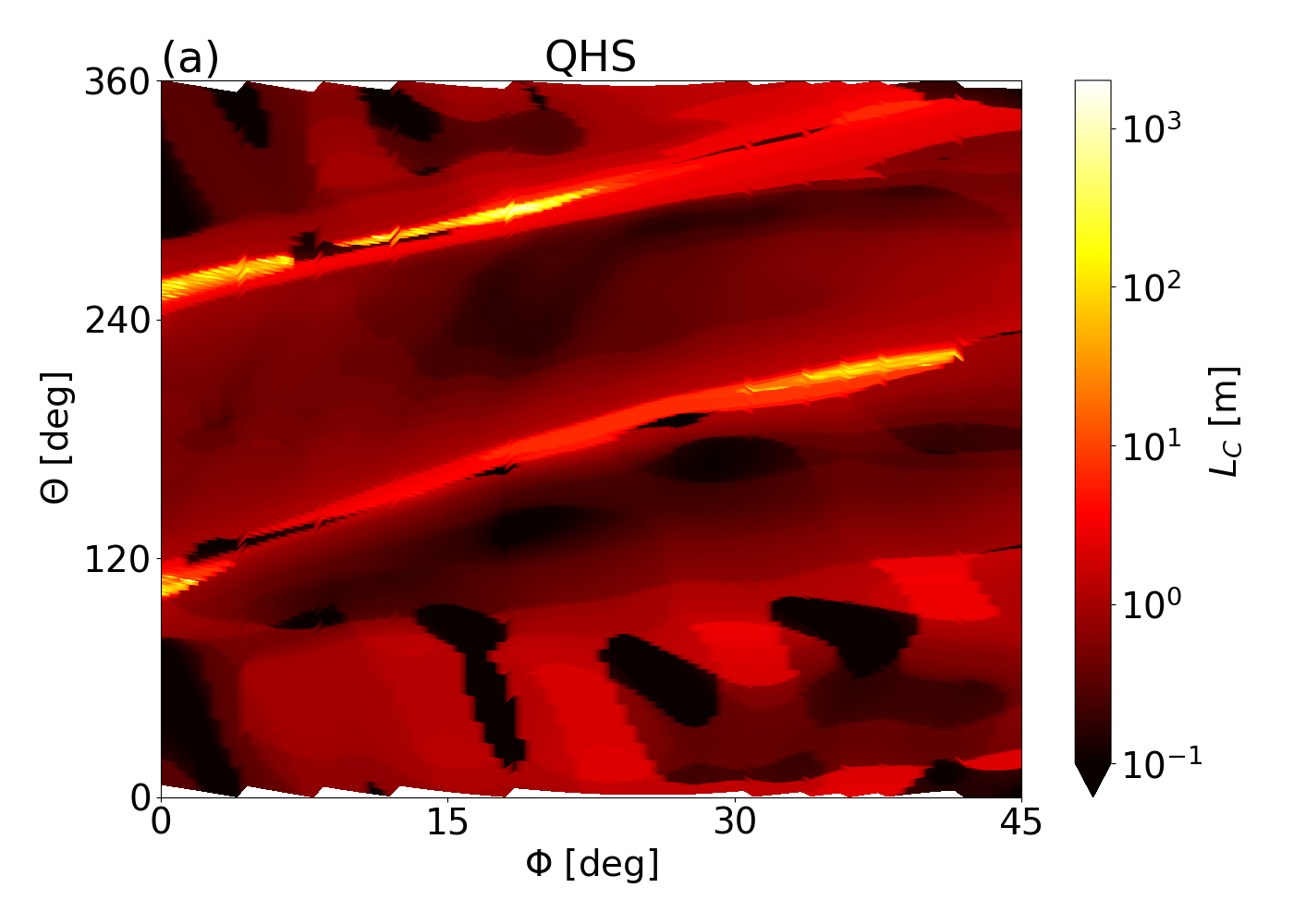}
    \includegraphics[scale=0.182]{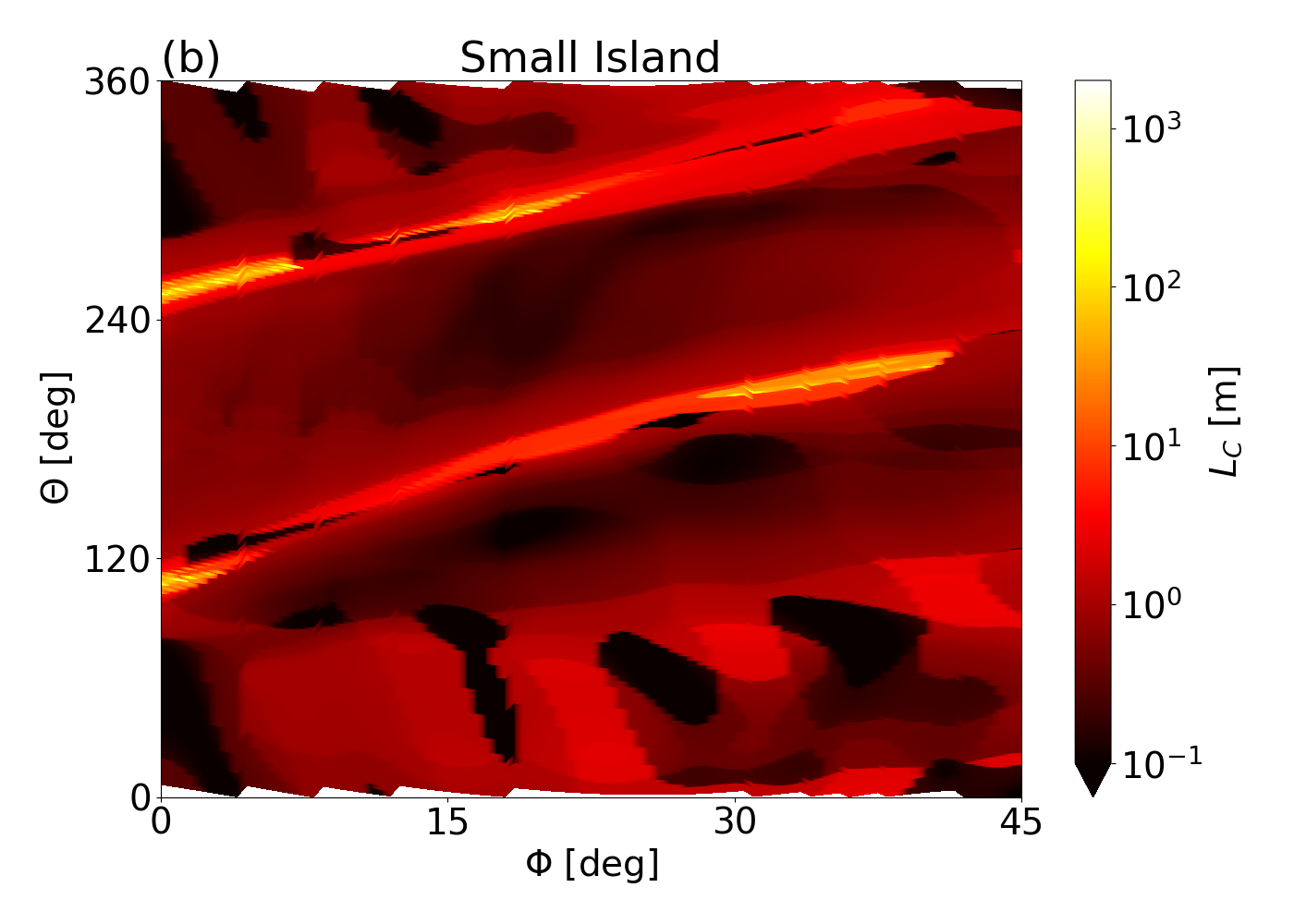}
    \includegraphics[scale=0.182]{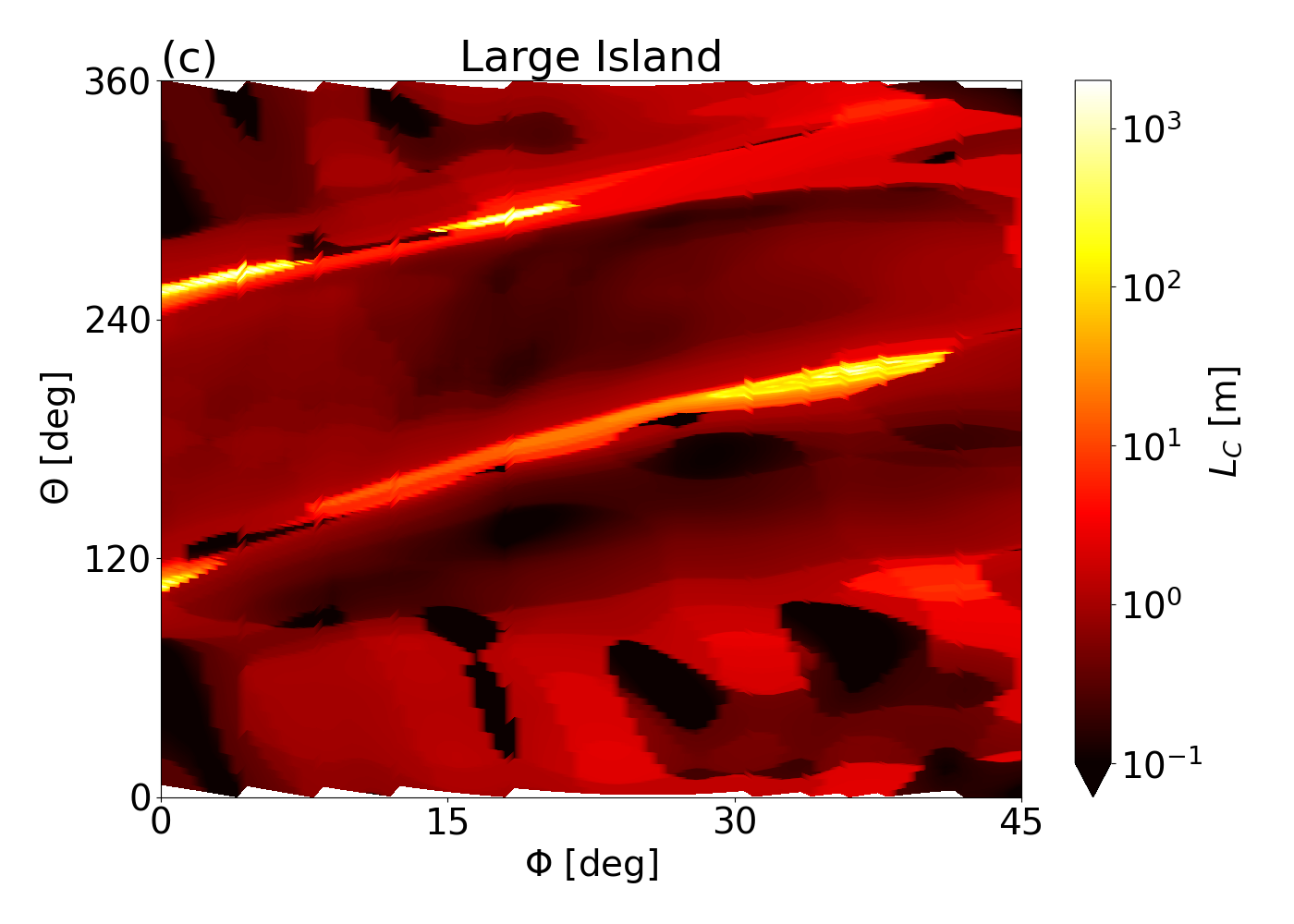}
    \includegraphics[scale=0.182]{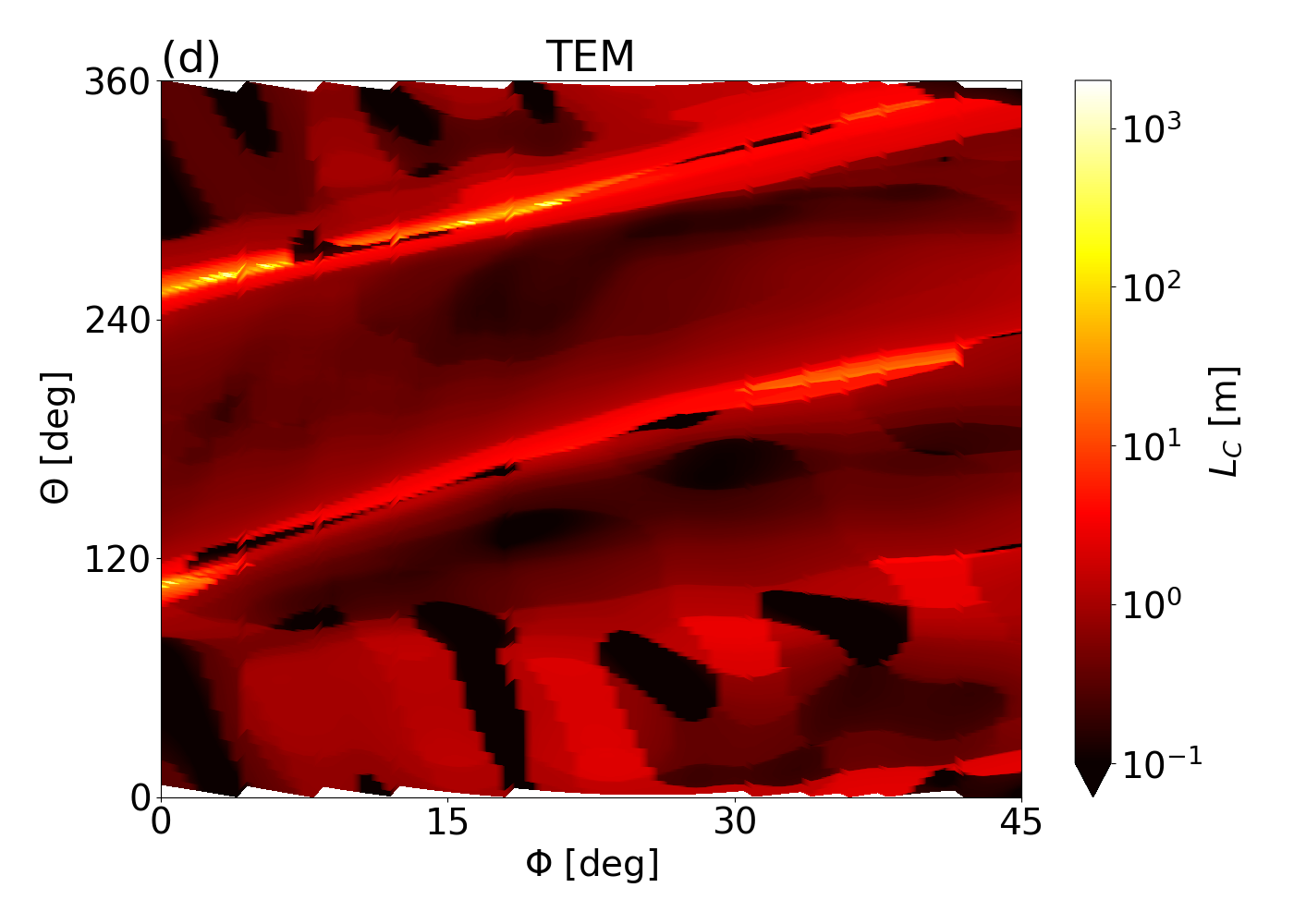}
    \caption{Magnetic footprint $\sim 0.5 \ $cm away from the expanded vessel wall for (a) QHS, (b) small island, (c) large island, and (d) TEM.}
    \label{fig:Lc_target}
\end{figure}

Further examination of the plasma wall behavior is performed by examining the edge field line interaction near the wall. In figures \ref{fig:Lc_target}-\ref{fig:high_res}, a dense mesh approximately $0.5 \ $cm away from the lofted vessel wall was generated to sample the field lines in the vicinity of the wall, similar to what was done for CTH in \cite{garcia_exploration_2023}. HSX has 4-field symmetry and previous work shows that the field line behavior for each half-field period is mirrored for the next half-field period \cite{bader_hsx_2017}. Therefore, we simulate the first half-field period in order to examine the details of the PWI.

Figure \ref{fig:Lc_target} plots $L_C(\phi,\theta)$ to show the magnetic footprint expected on the wall. The magnetic footprint is calculated by following field lines near the wall for $1 \ $km or until they strike the wall. The connection length of these field lines is shown as a contour map on the sampled mesh near the PFC. Magnetic footprints in general serve as a proxy for heat and particle flux deposition on the PFC, where long $L_C$ is directly related to the heat flux \cite{abdullaev_stable_2014,garcia_exploration_2023}. The field line intersection pattern on the mesh confirms the resilient strike line feature of HSX seen in previous work \cite{bader_hsx_2017} where we note that reference \cite{bader_hsx_2017} did not observe resiliency for the large island case. This will be discussed later. The resilient feature is observed by the long $L_C \sim 1 \ $km helical band corresponding to white and yellow on the colorbar. These helical bands of long $L_C$ are approximately in the regions of $0^{\circ} \leq \phi \leq 20^{\circ}$ near $\theta \sim 240^{\circ}$ along with $30^{\circ} \leq \phi \leq 40^{\circ}$ near $\theta \sim 200^{\circ}$. The red and black regions of figure \ref{fig:Lc_target} with $L_C \leq 10 \ $m represent field lines of the far scrape-off layer (SOL). As these field lines are not contributing to the resilient helical band, their PWI will not be examined further. The most apparent difference can be seen between the large island result (figure \ref{fig:Lc_target} (c)) and the TEM result (figure \ref{fig:Lc_target} (d)). Here, the very bright region of very long connection length that is seen in the large island case is mostly missing in the TEM case. Nevertheless, it still manifests as a band of moderate (orange) connection length.

\subsection{Radial Connection from Lofted Wall}
\label{subsection:radpen}

Another figure of merit used in the tokamak community for characterization of resonant magnetic perturbations is the radial penetration of field lines \cite{frerichs_pattern_2015,frerichs_exploration_2016}. Typically, these calculations are done by evaluating the enclosed poloidal flux of the unperturbed magnetic configuration. Then, field lines are launched in the perturbed configuration while cataloguing the minimum value of poloidal flux they achieve. For the equilibria in this work, we attempt a similar calculation. However, there is no natural enclosed flux value that extends beyond the LCFS, therefore, we use a simplified approach that instead calculates the distance from the field line to the LCFS along its trajectory and catalogue the minimum value. We denote this as the radial connection of the field line. The calculation is further complicated in that the LCFS differs significantly for each configuration. We choose to calculate each result with respect to the LCFS of the same configuration rather than using a standard case. While this complicates comparisons across the configurations, it is necessary due to the significant LCFS differences.

\begin{figure}[H]
    \centering
    \includegraphics[scale=0.182]{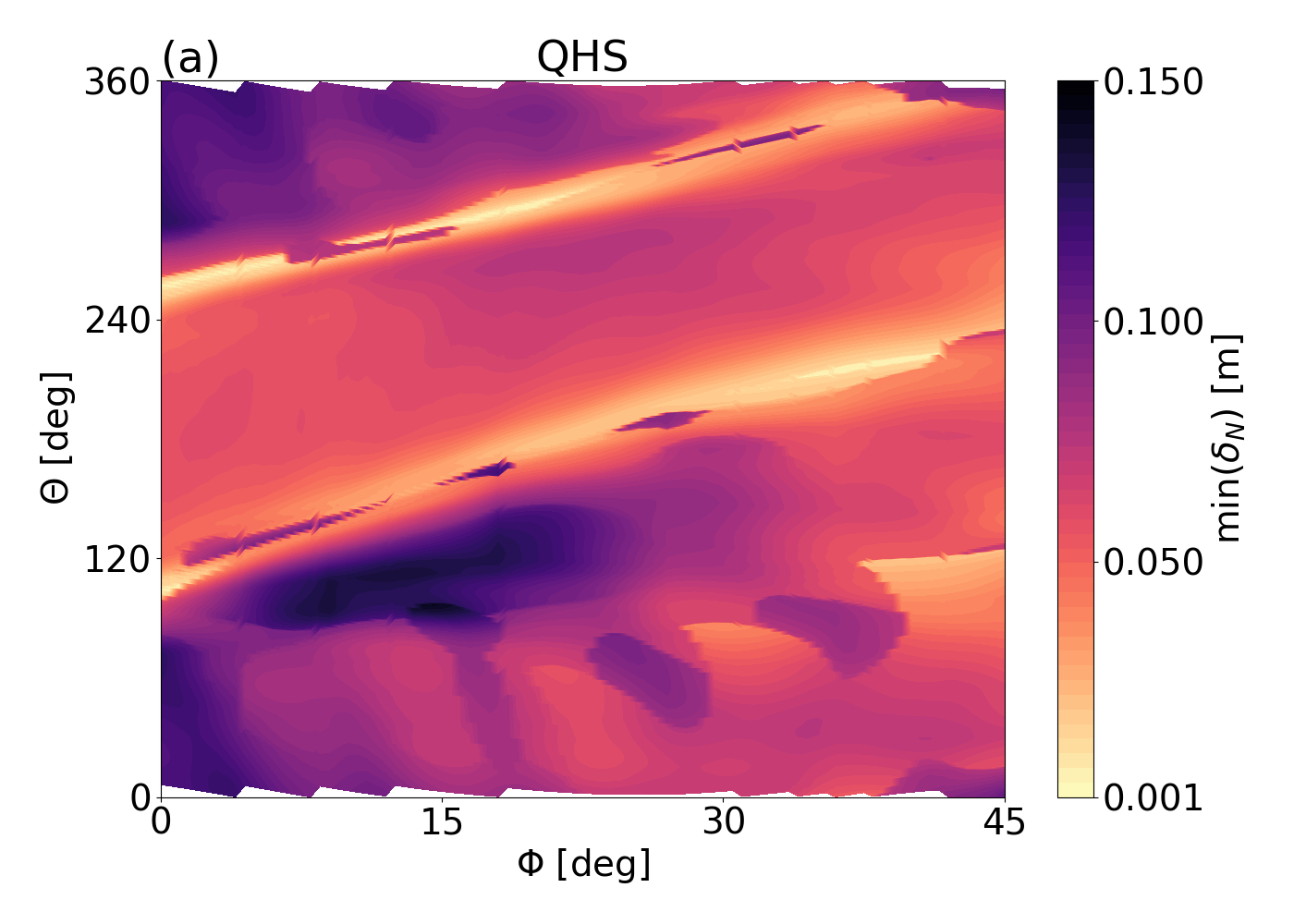}
    \includegraphics[scale=0.182]{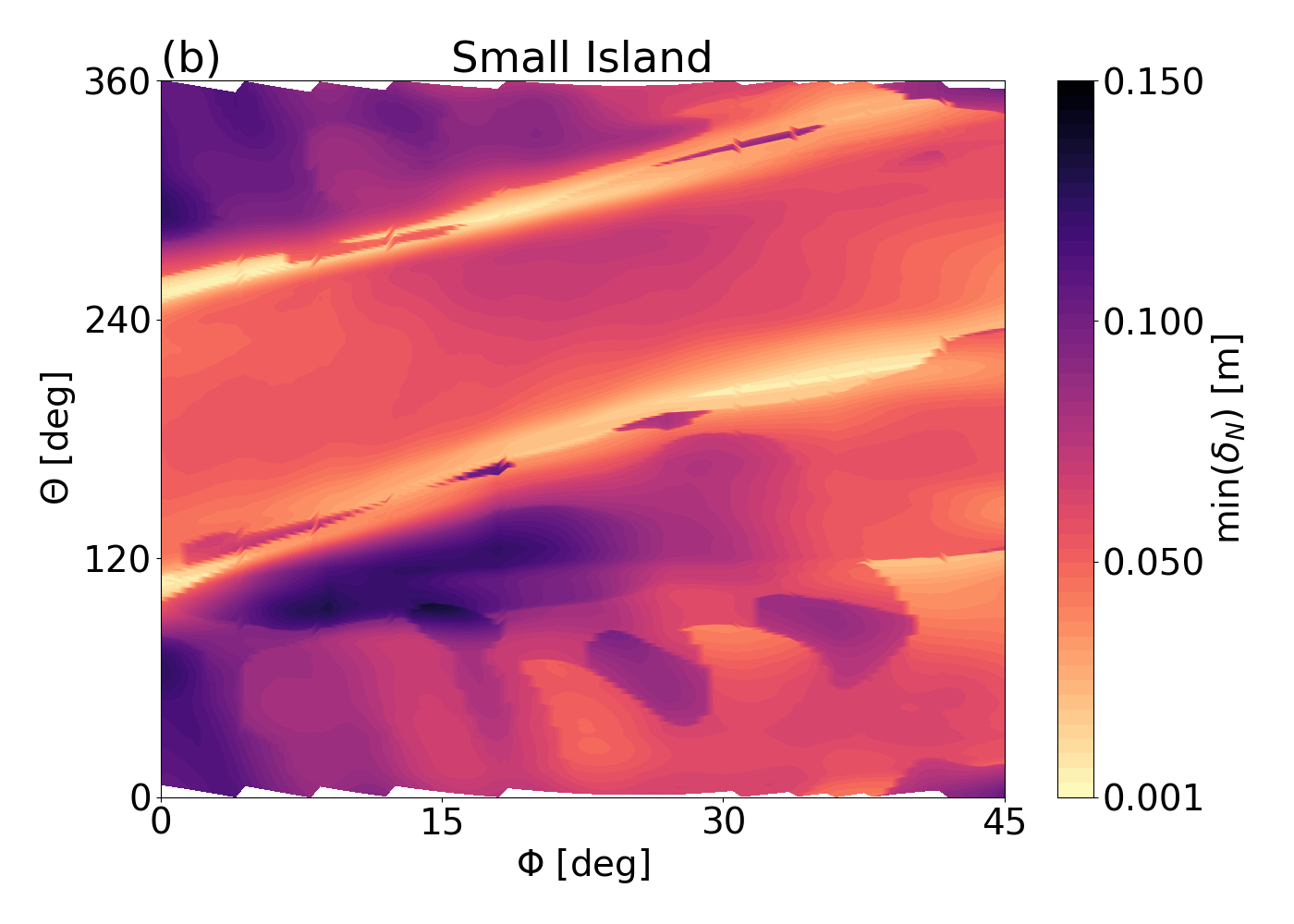}
    \includegraphics[scale=0.182]{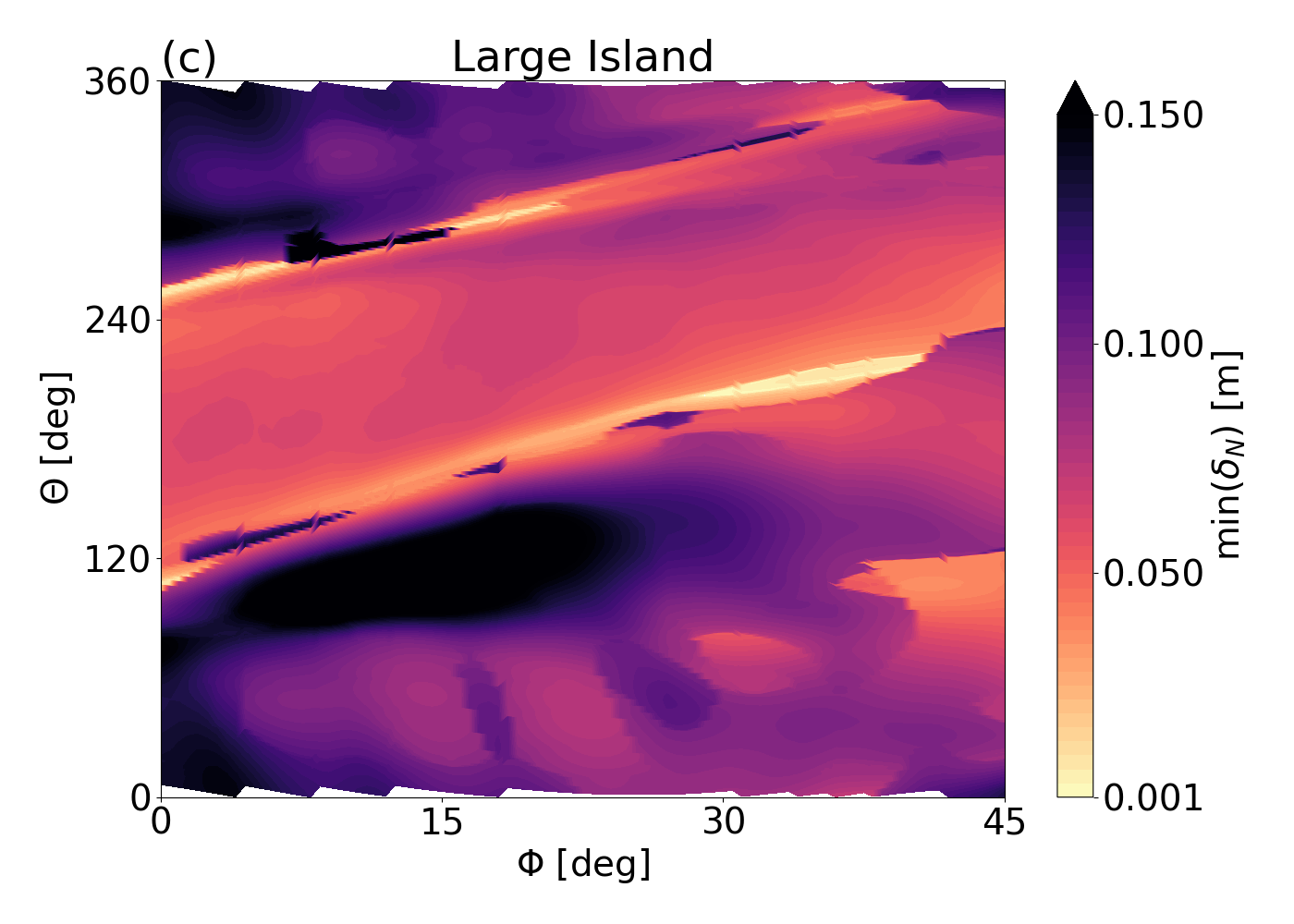}
    \includegraphics[scale=0.182]{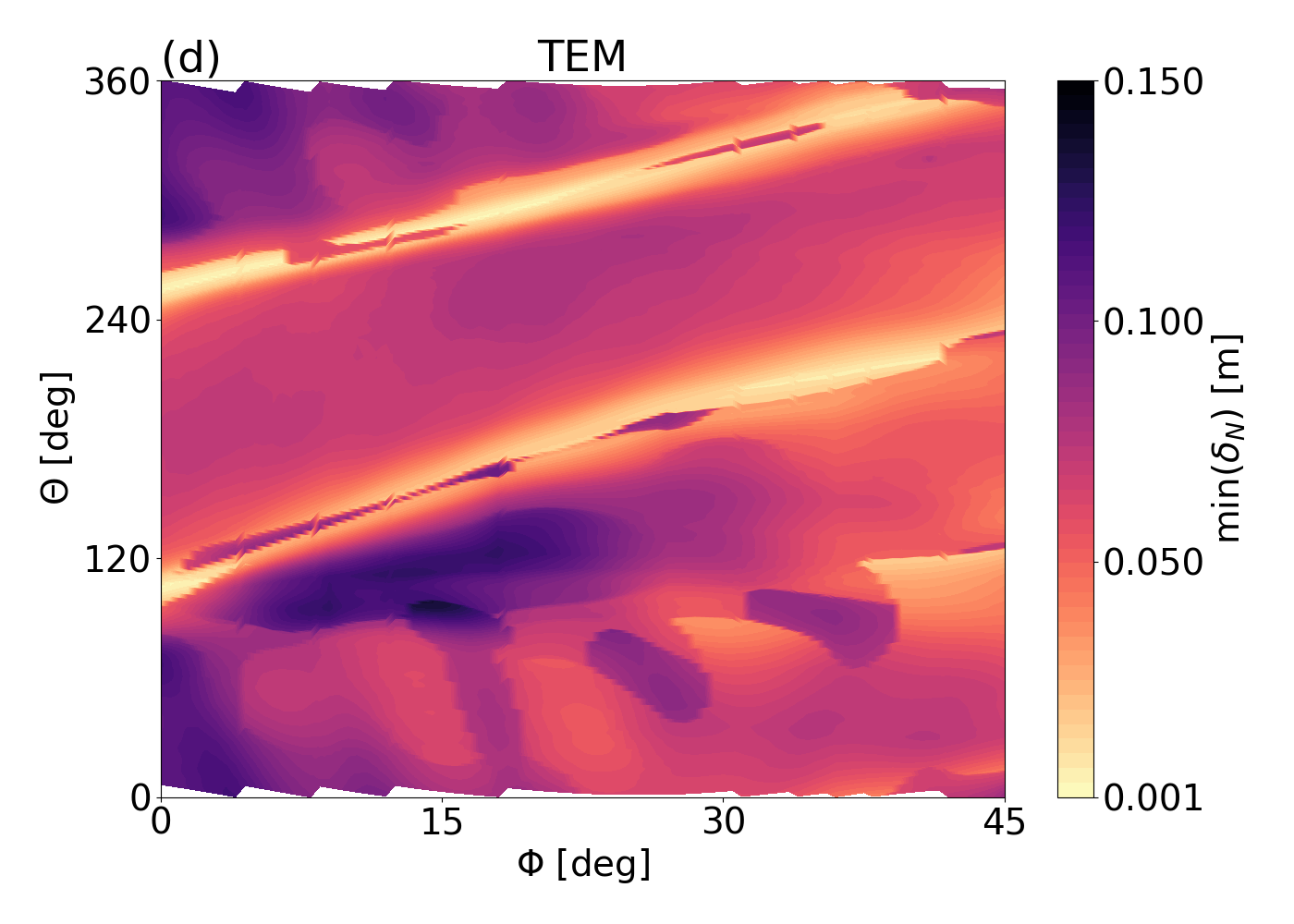}
    \caption{Radial connection $\sim 0.5 \ $cm away from the expanded vessel wall for (a) QHS, (b) small island, (c) large island, and (d) TEM.}
    \label{fig:PsiN_target}
\end{figure}

Using the same mesh for the magnetic footprint computation, the minimum radial connection min$(\delta_N)$ of the field lines is simulated and plotted in figure \ref{fig:PsiN_target}. Similar to the magnetic footprint calculation, the field lines are followed for a maximum of $L_C= 1 \ $km in both directions. A positive min$(\delta_N)$ indicates that this distance is radially outward. Like figure \ref{fig:Lc_target}, the radial connection calculated on the mesh is very qualitatively similar across the 4 cases despite the use of different LCFS values for all four configurations. Furthermore, the PWI pattern of figure \ref{fig:PsiN_target} also highly resembles the magnetic footprint pattern of figure \ref{fig:Lc_target}. In the regions of long $L_C$ in figure \ref{fig:Lc_target}, the distance from the LCFS is minimal and corresponds to min$(\delta_N) \sim 0.1 \ $cm. This calculation combined with the magnetic footprint of figure \ref{fig:Lc_target} demonstrates that the resilient feature (regions of high $L_C$) seen across all configurations is dominated by long connection length field lines that approach very closely to the LCFS. These are the field lines that most influence the deposition of particles and energy onto the PFCs. This qualitative agreement can also be seen in figure \ref{fig:high_res} which shows min$(\delta_N)$ and $L_C$ on a higher resolution mesh for a smaller poloidal and toroidal range which will be discussed further in the next section. 

From these figures, we observe that the deposition of particles and energy onto the PFC is connected to the plasma core via very long $L_C$ with a short radial distance from the LCFS. The next section is dedicated to using $L_C$ and min$(\delta_N)$ together to quantify the differences within the details of the PWI due to the different topological structures present across the edge configurations. 

\section{Radial Connection as a Metric for Determining Island Behavior}
\label{section:differences}
\begin{figure}[H]
    \centering
    \includegraphics[scale=0.182]{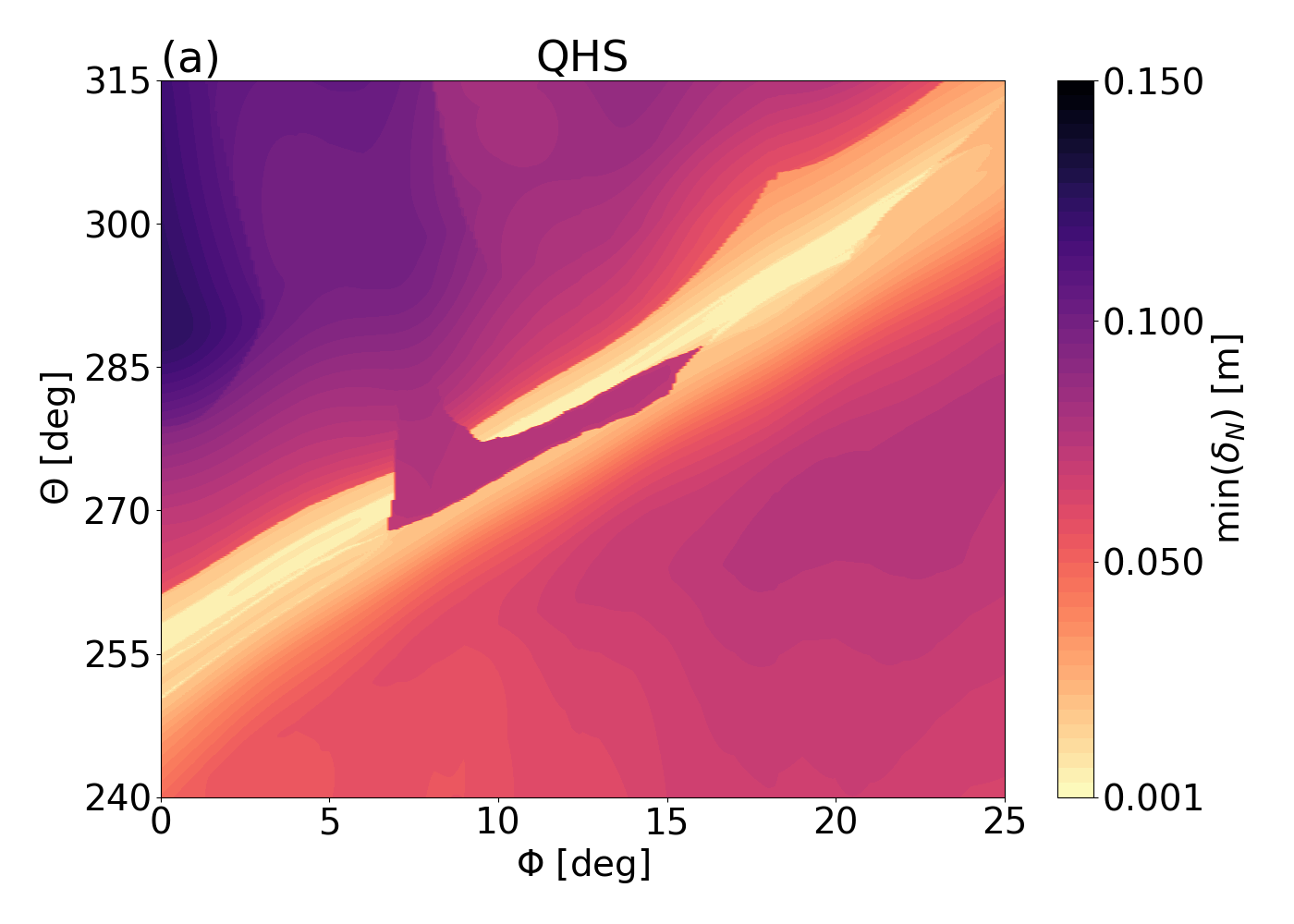}
    \includegraphics[scale=0.182]{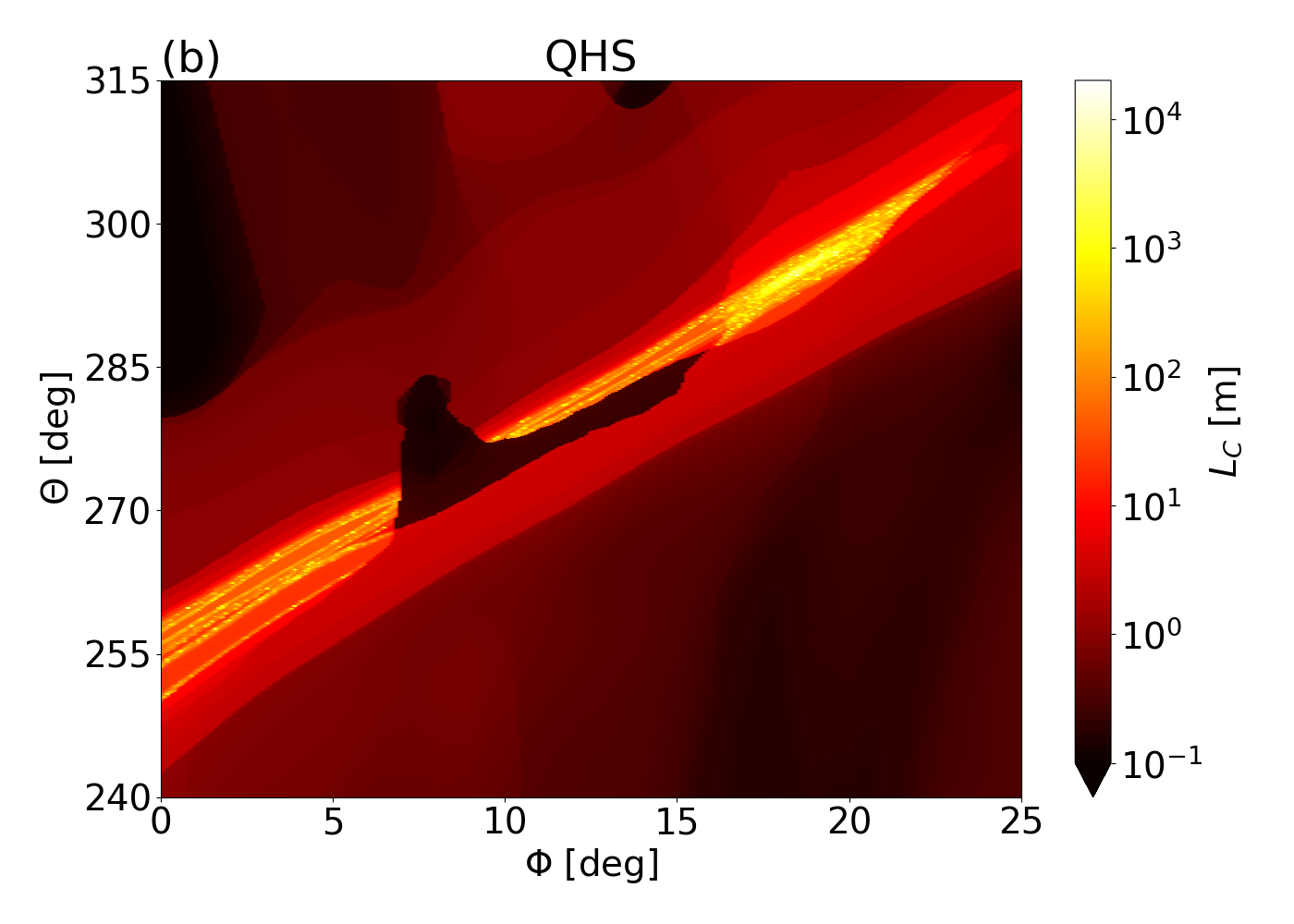}
    \includegraphics[scale=0.182]{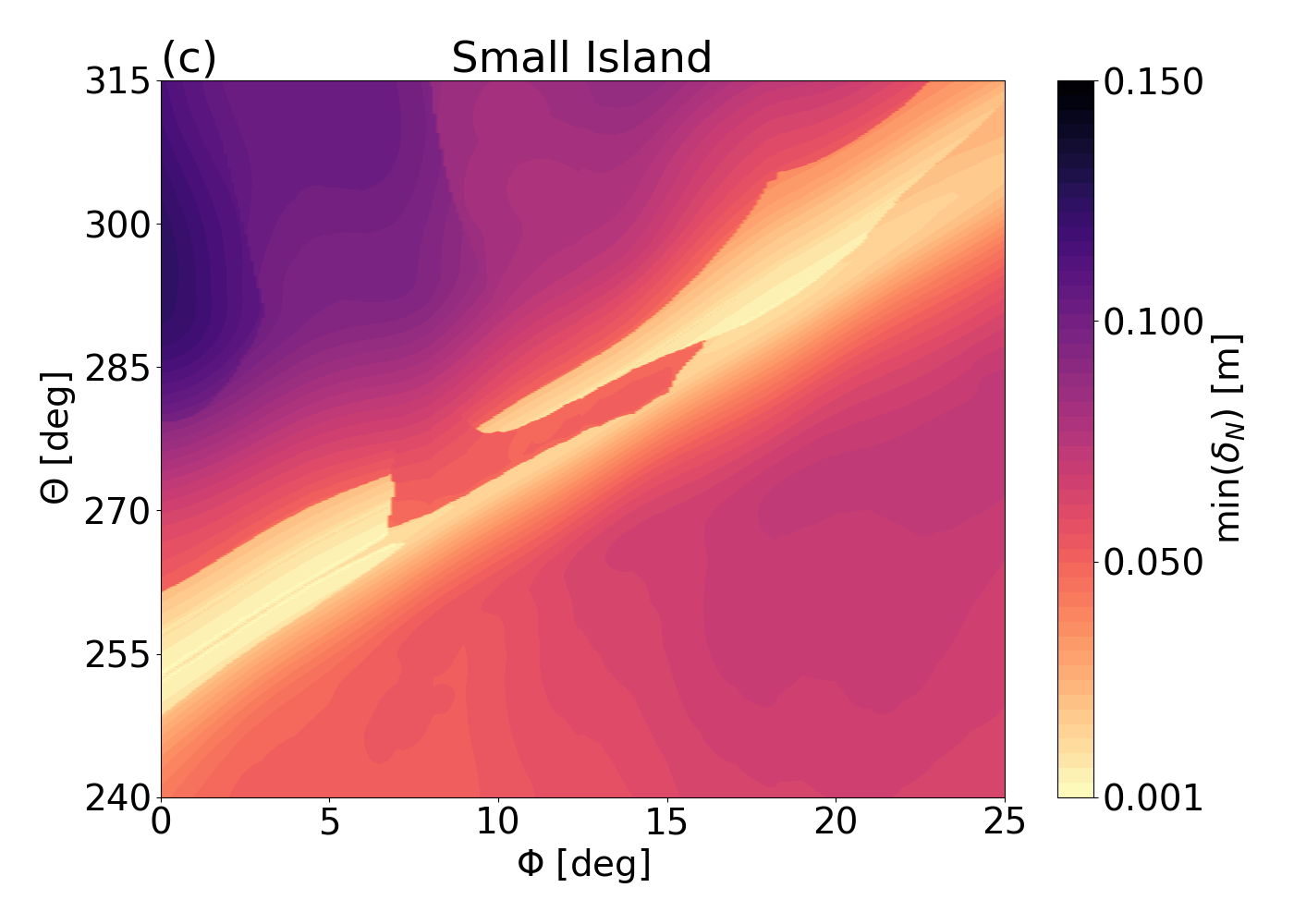} 
    \includegraphics[scale=0.182]{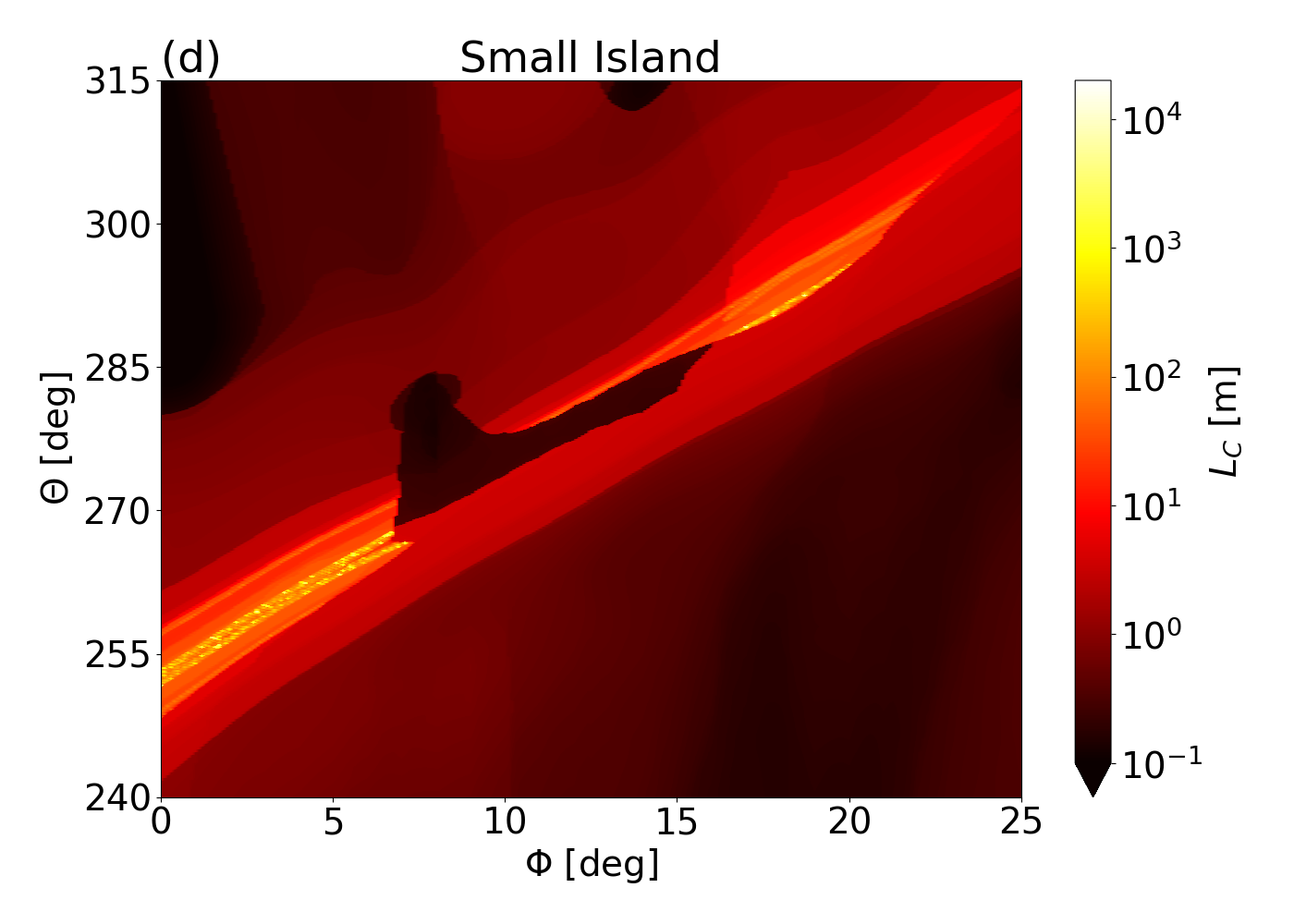}
    \includegraphics[scale=0.182]{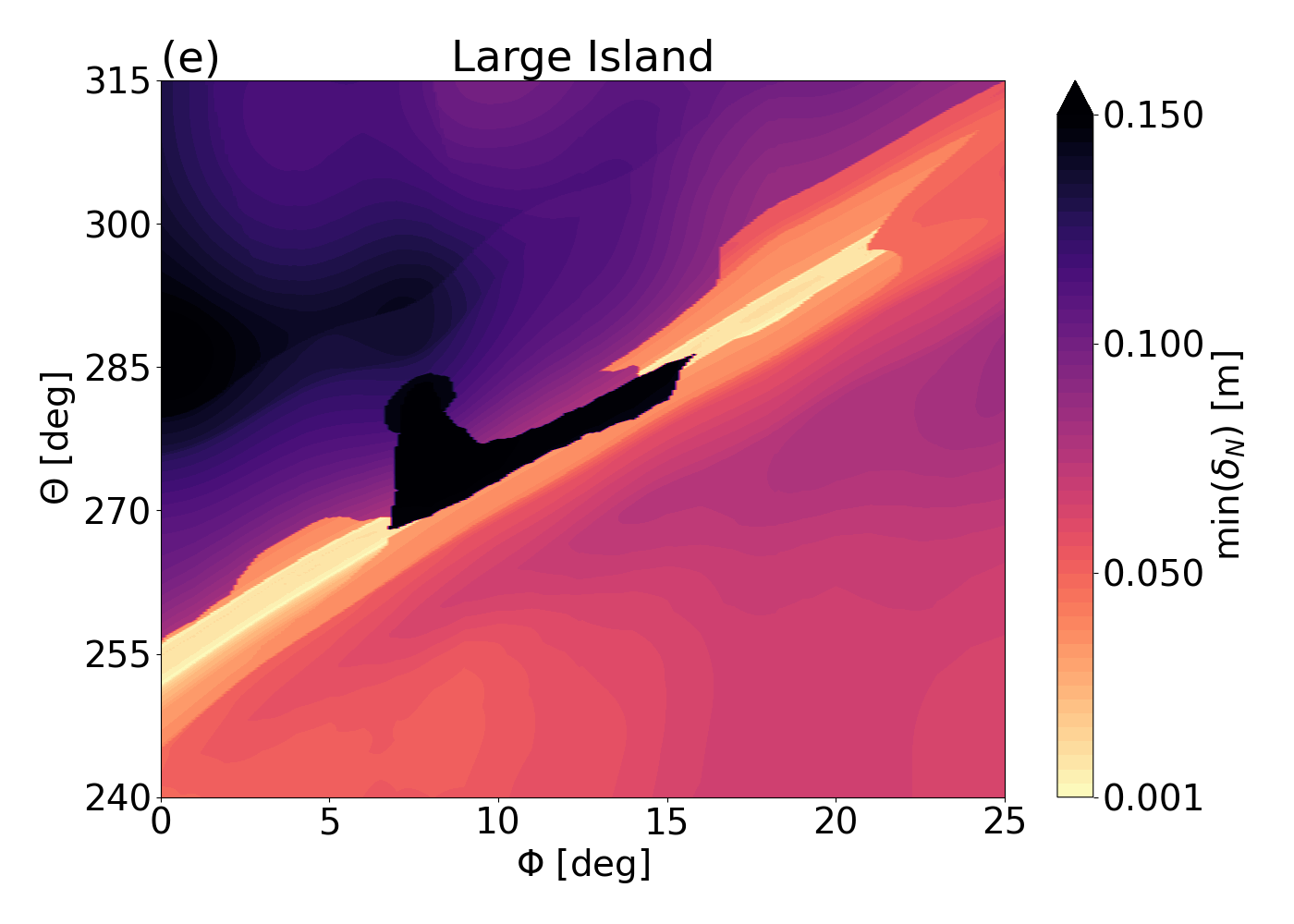}
    \includegraphics[scale=0.182]{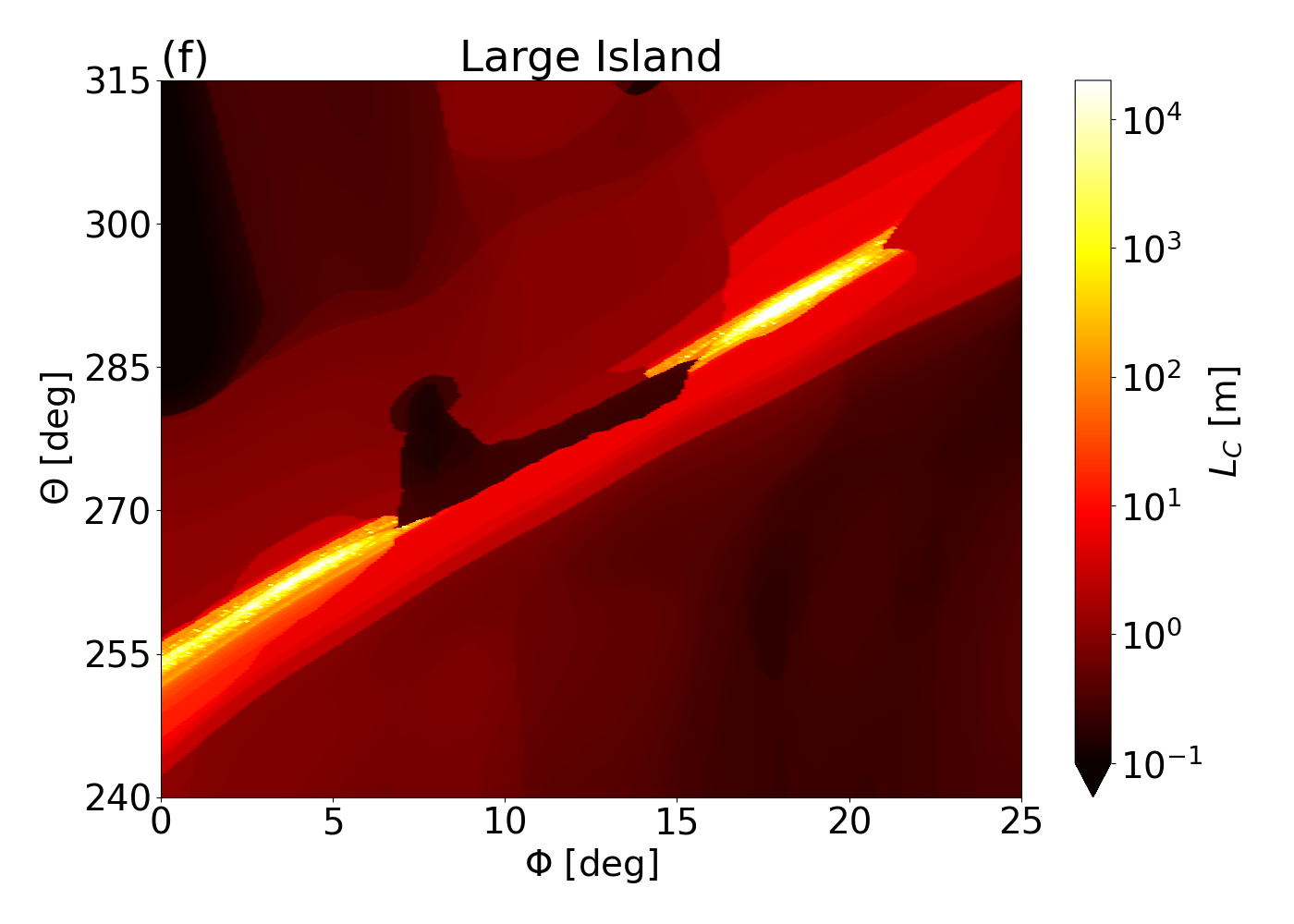}
    \includegraphics[scale=0.182]{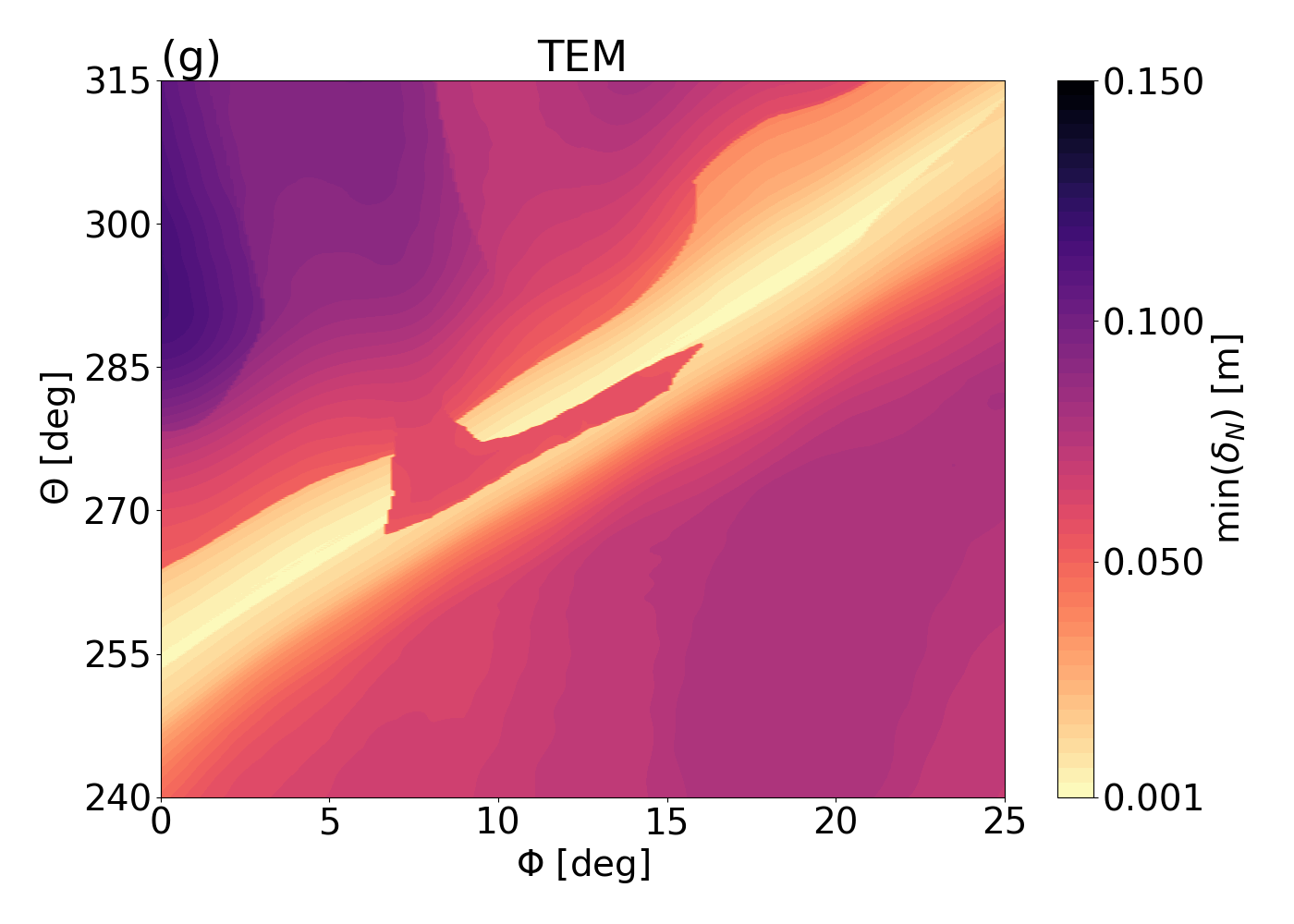}
    \includegraphics[scale=0.182]{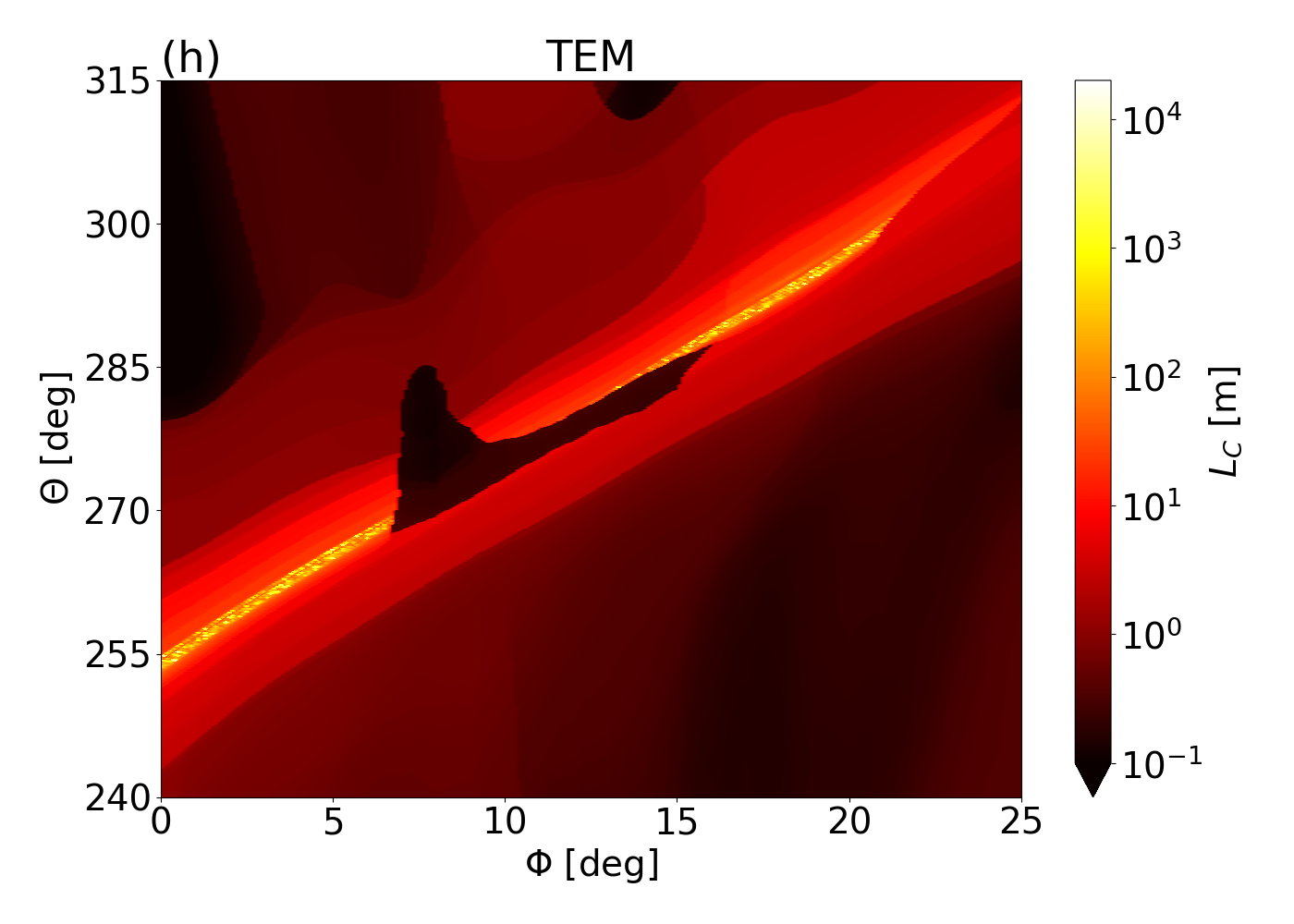} 
    \caption{High resolution comparison of min$(\delta_N)$ (left column) and $L_C$ (right column).}
    \label{fig:high_res}
\end{figure}

\begin{figure}[H]
    \centering
    \includegraphics[scale=0.65]{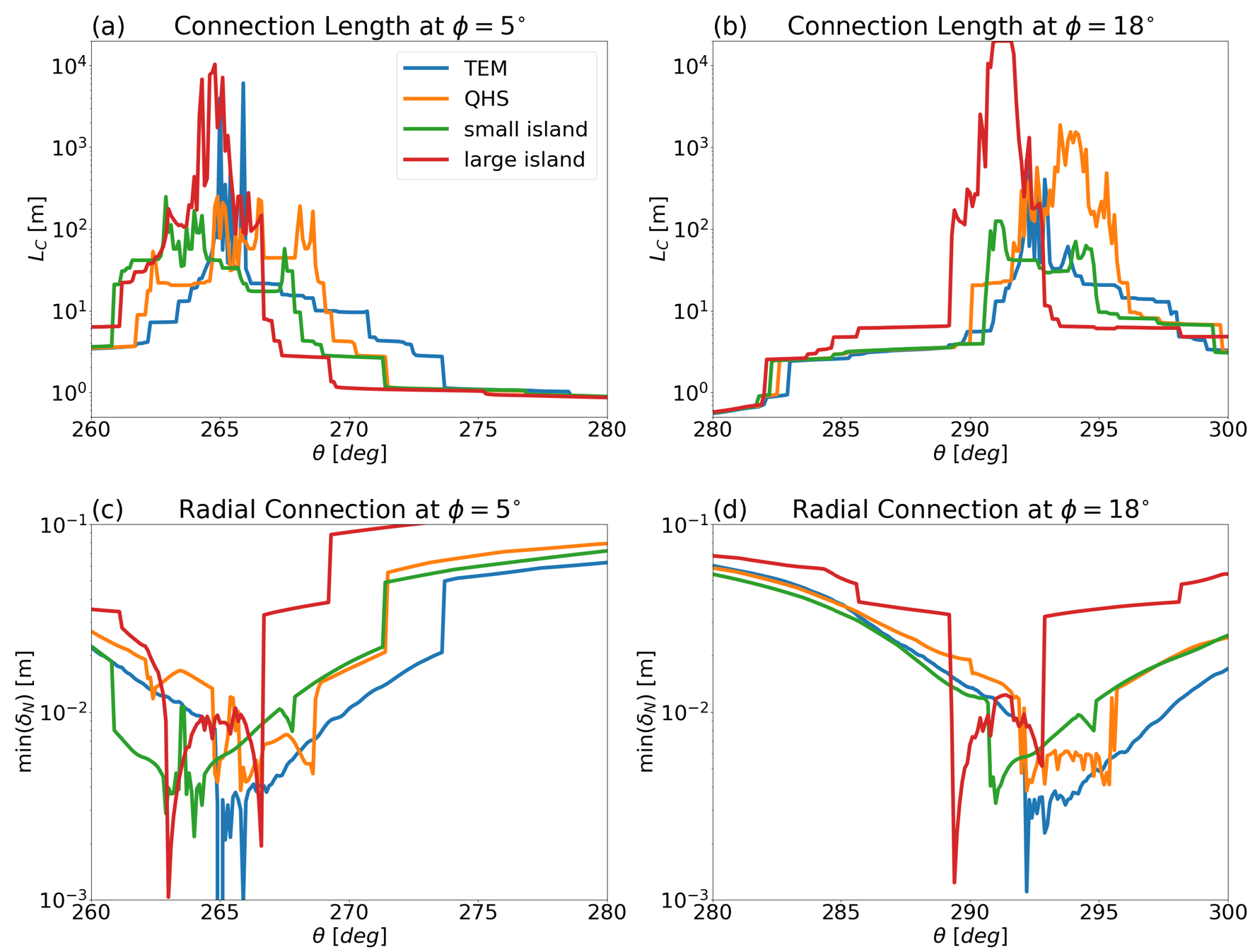}
    \caption{Plots (a) and (b) of $L_C$ as a function of $\theta$ at $\phi=5^\circ$ and $\phi=18^\circ$, respectively. Plots (c) and (d) of min$(\delta_N)$ as a function of $\theta$ at $\phi=5^\circ$ and $\phi=18^\circ$, respectively.}
    \label{fig:2D_zoom_log_10^4_higherres}
\end{figure}

\begin{figure}[H]
    \centering
    \includegraphics[scale=0.5]{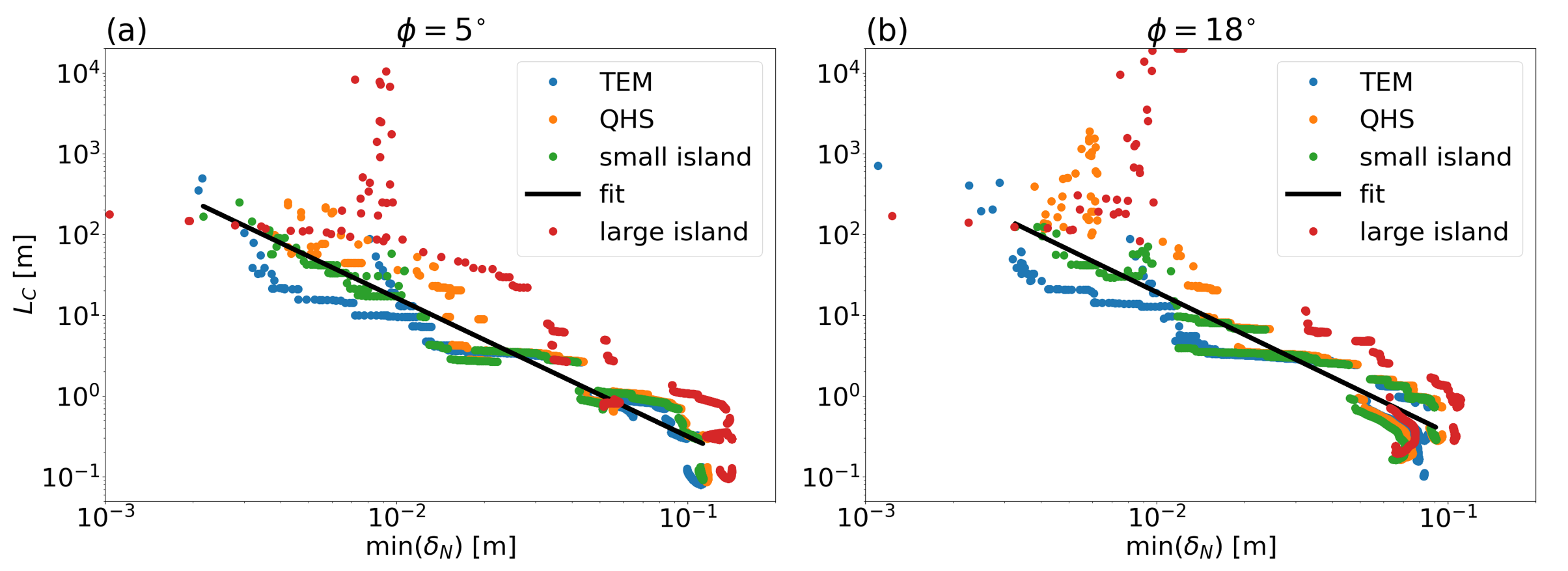}
    \caption{$L_C$ vs min$(\delta_N)$ for (a) $\phi=5^\circ$ (from figure \ref{fig:2D_zoom_log_10^4_higherres} (a) and (c)) and (b) $\phi = 18^\circ$ (from figure \ref{fig:2D_zoom_log_10^4_higherres} (b) and (d)).}
    \label{fig:Lc_vs_DeltaN}
\end{figure}

\begin{figure}[H]
    \centering
    \includegraphics[scale=0.21]{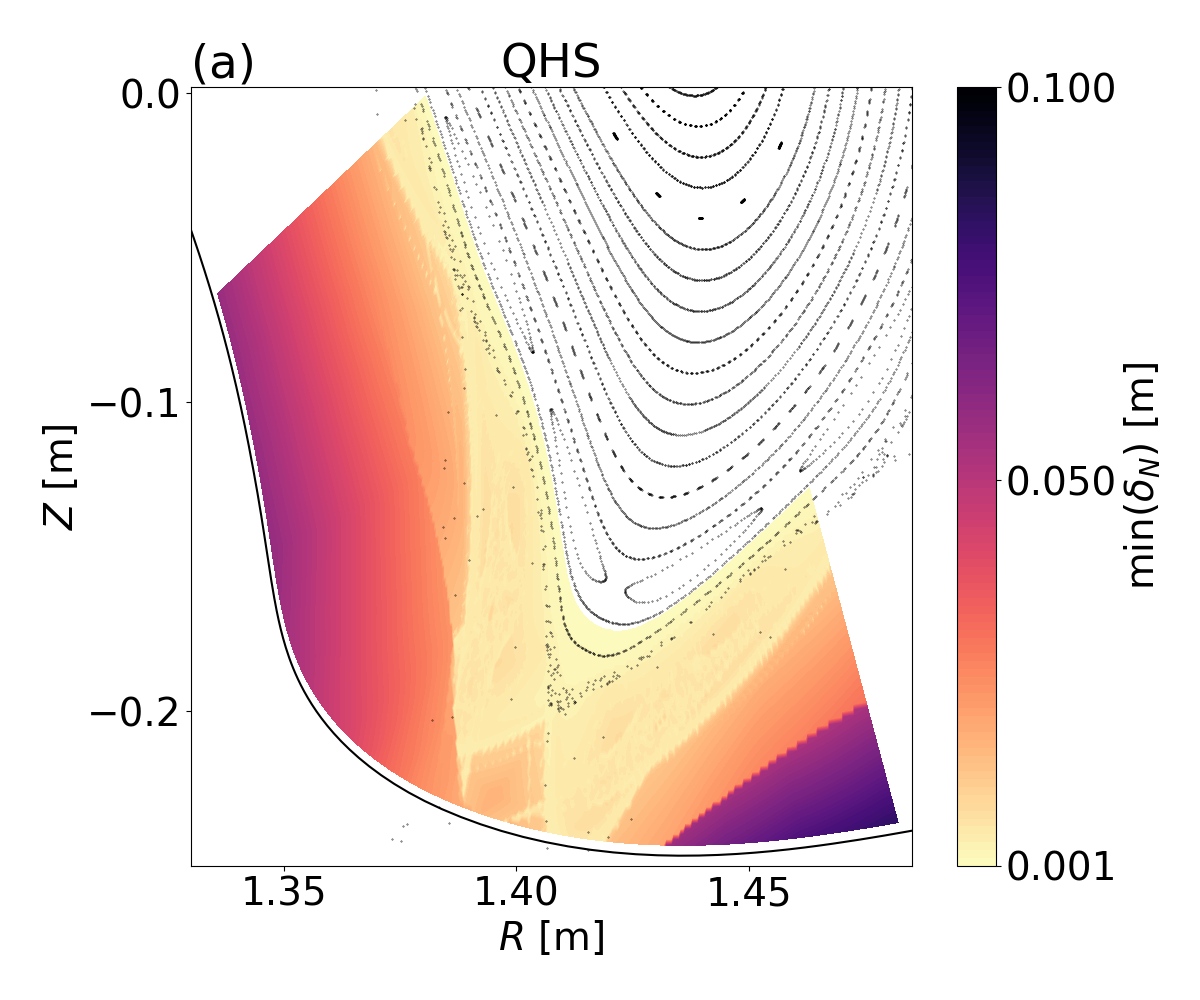}
    \includegraphics[scale=0.21]{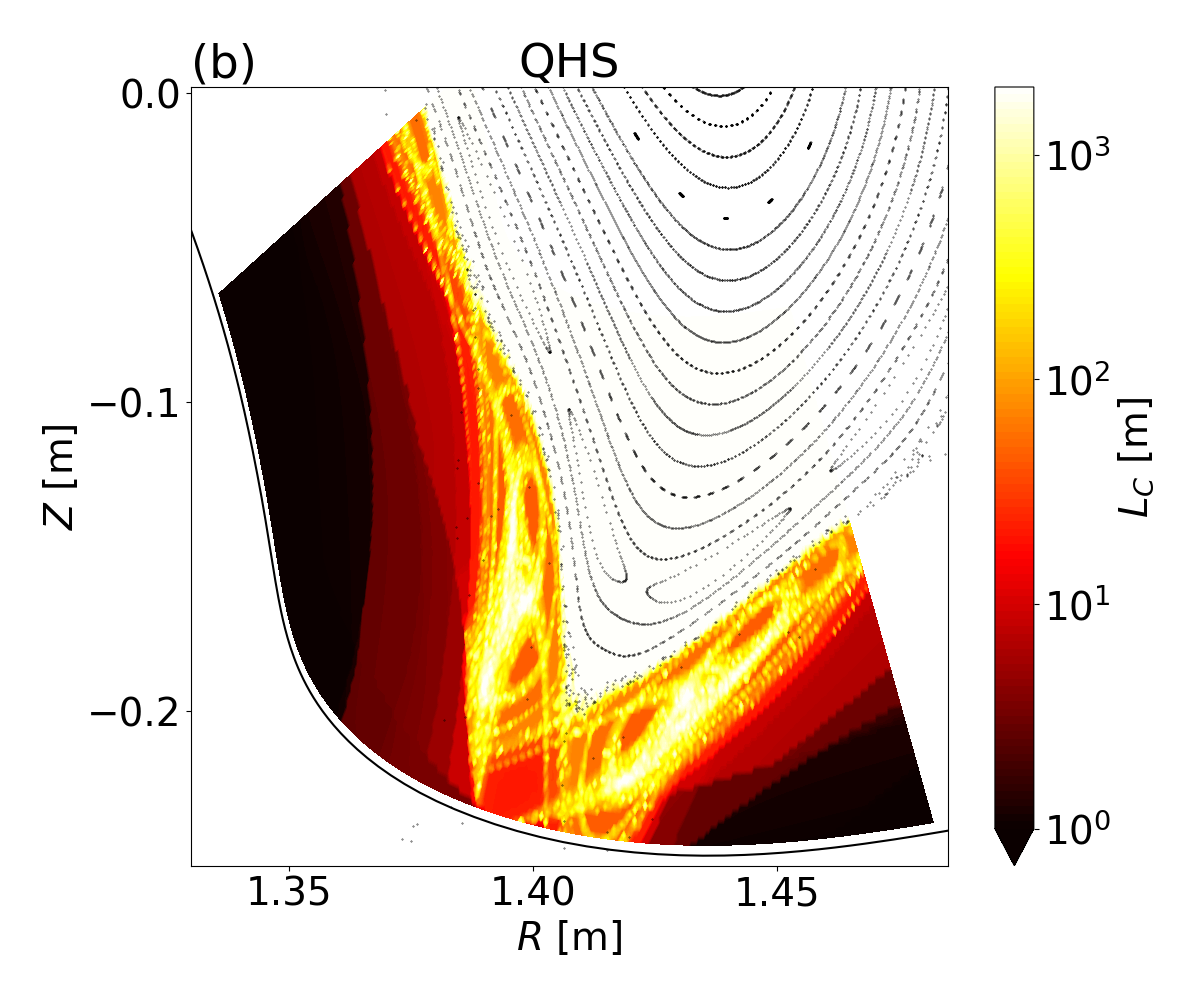}
    \includegraphics[scale=0.21]{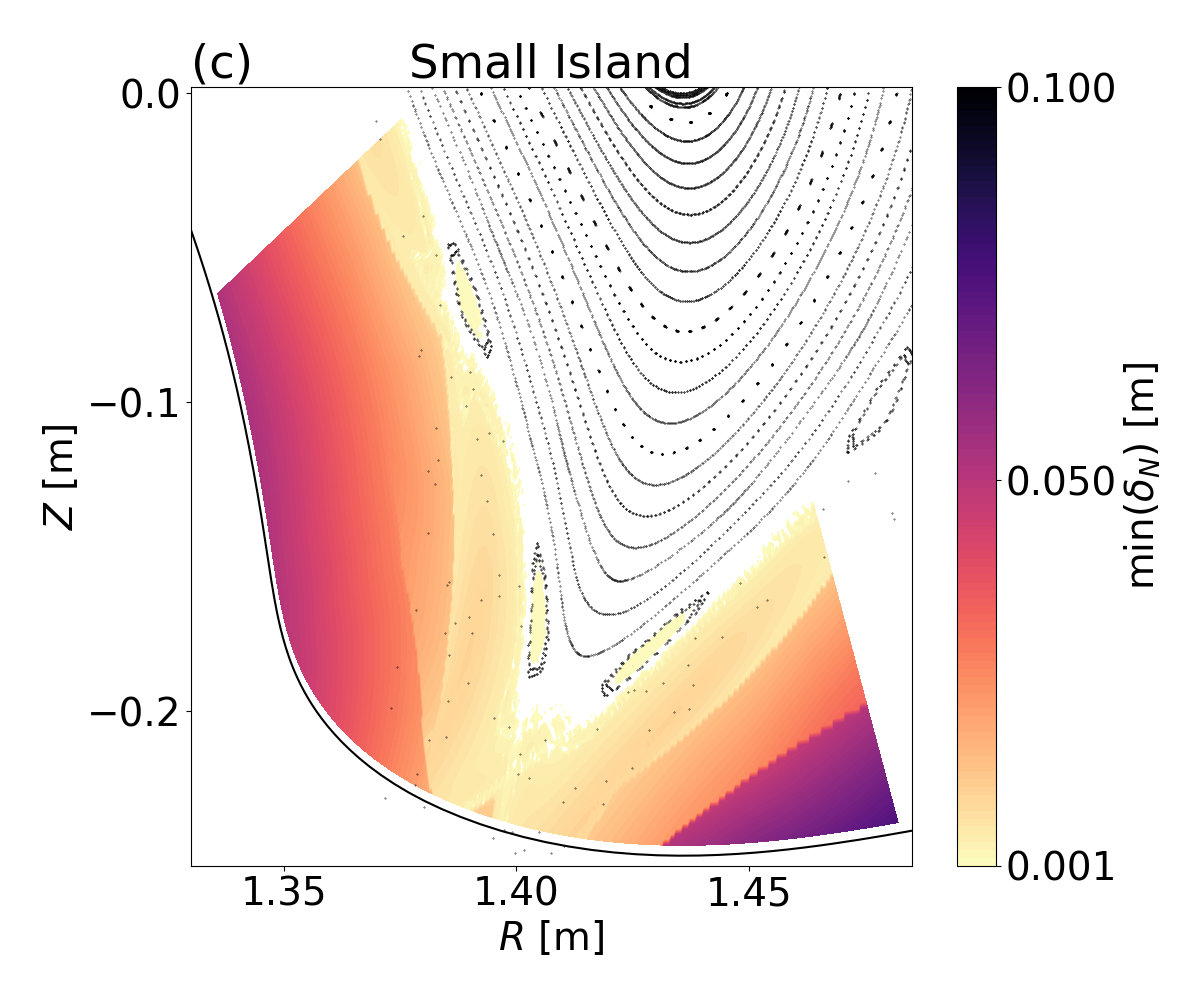}
    \includegraphics[scale=0.21]{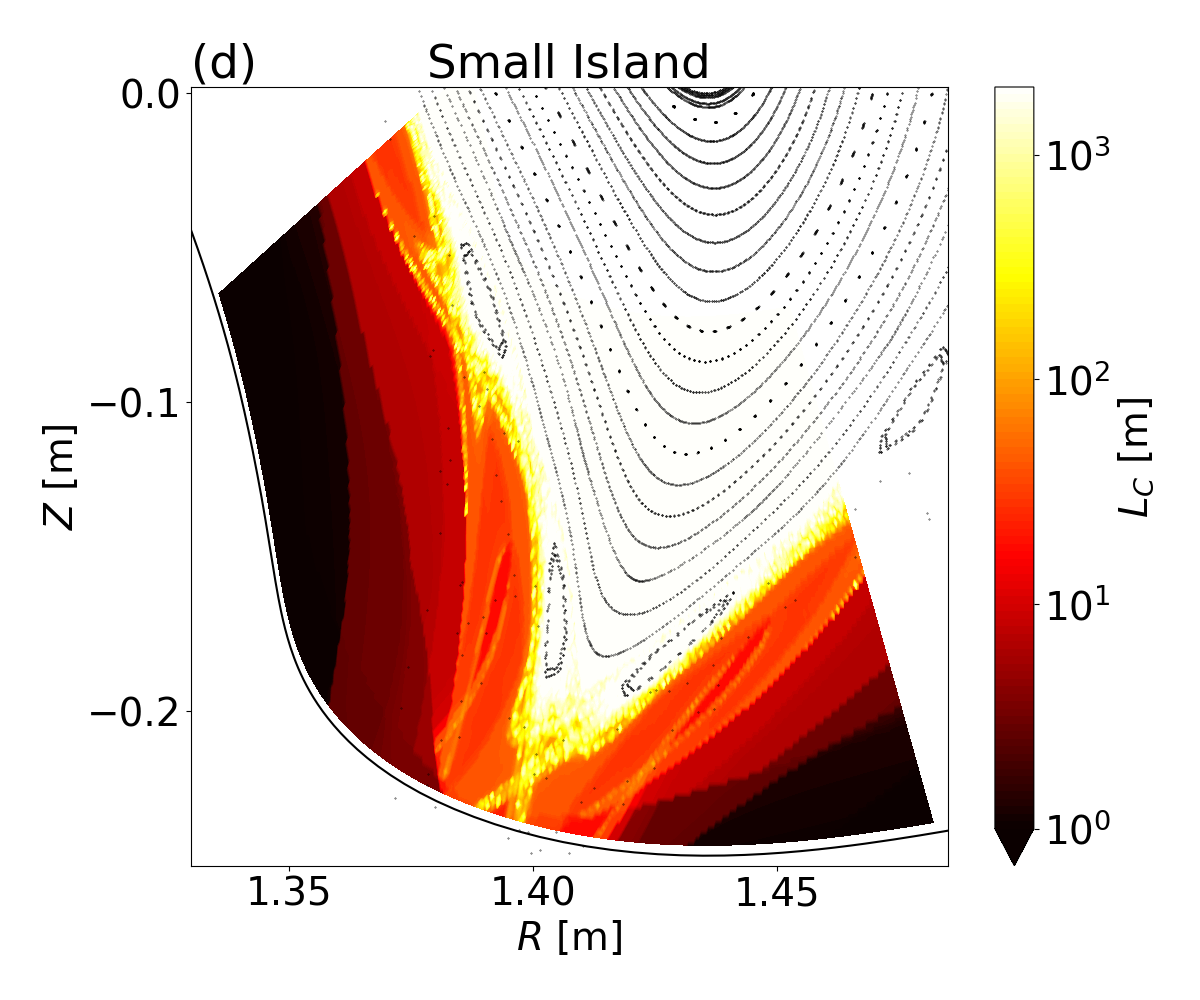}
    \includegraphics[scale=0.21]{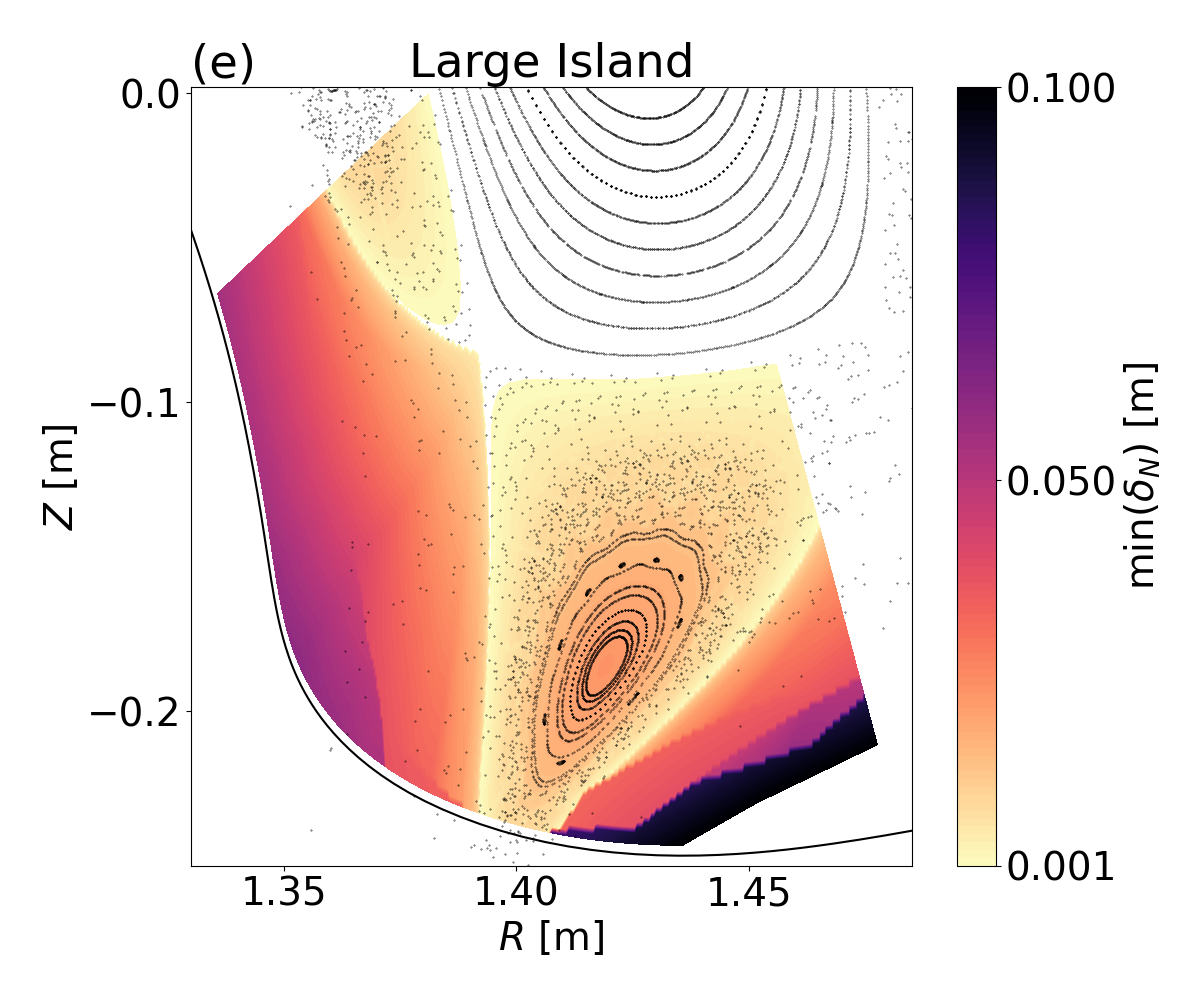} 
    \includegraphics[scale=0.21]{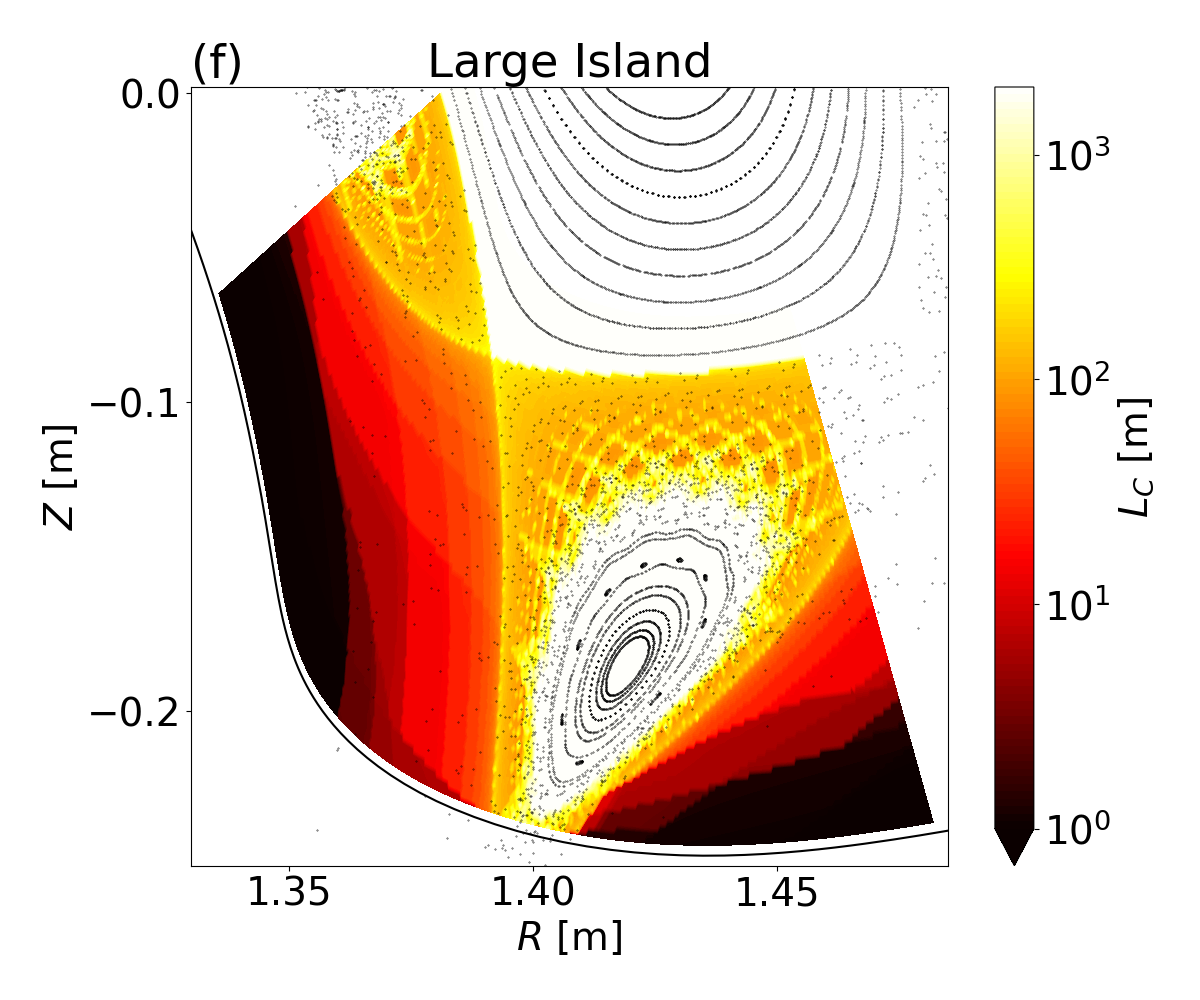}
    \includegraphics[scale=0.21]{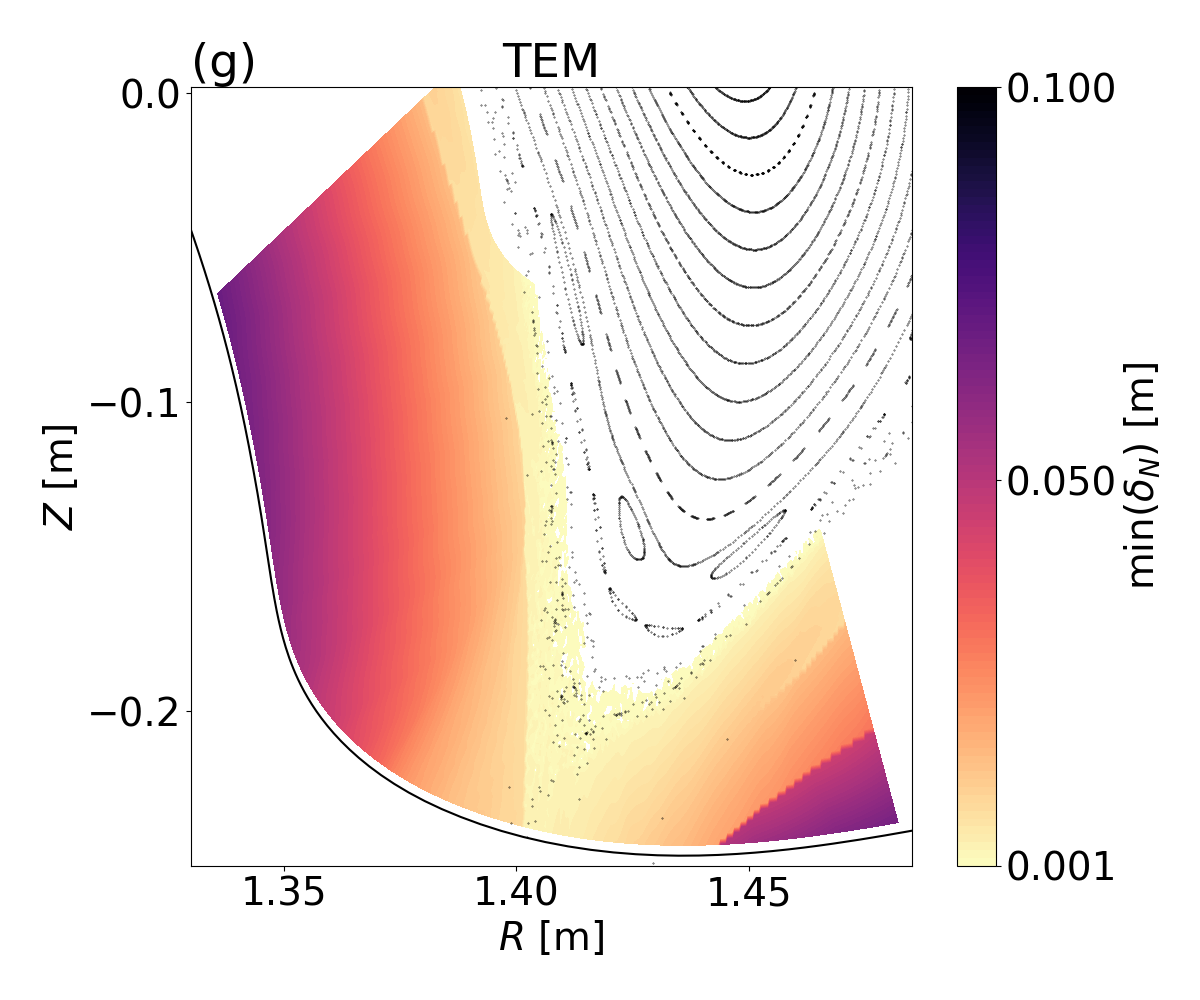}
    \includegraphics[scale=0.21]{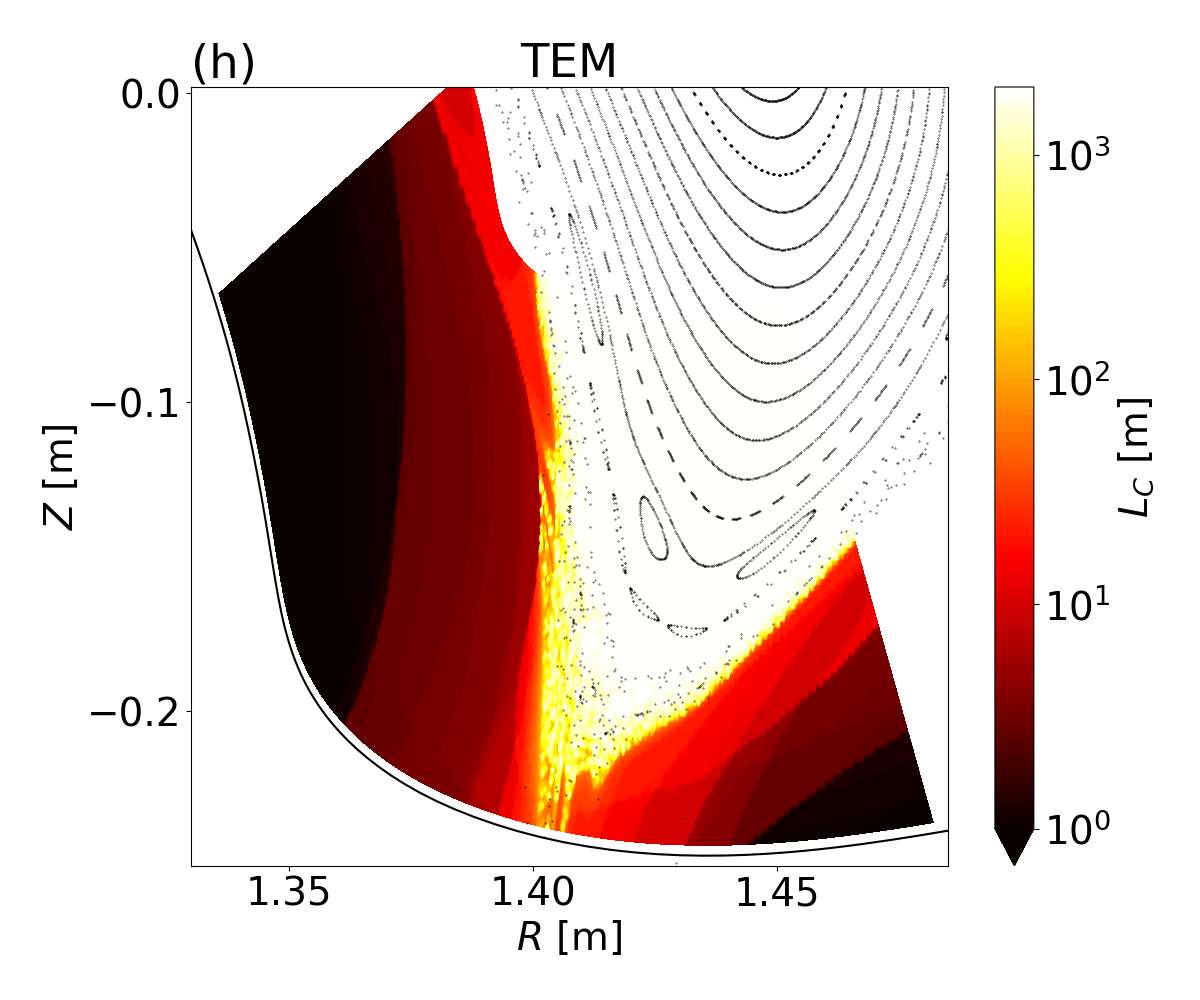}
    \caption{High resolution figures of the PWI corresponding to $\phi=5^\circ$ for each magnetic configuration. The left column is the radial connection and the right column is the connection length.}
    \label{fig:poincare5_Lc_DeltaN}
\end{figure}

\begin{figure}[H]
    \centering
    \includegraphics[scale=0.21]{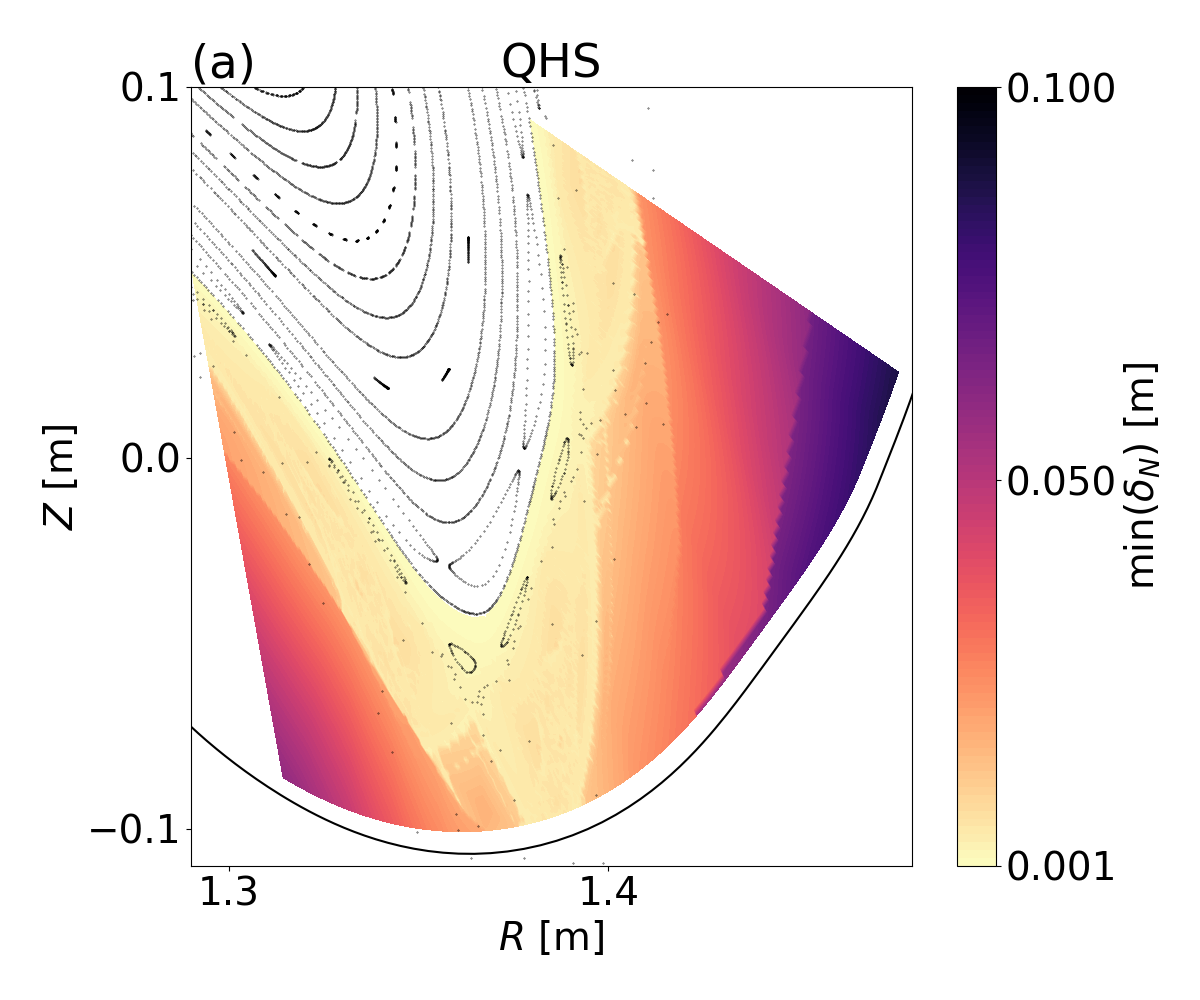}
    \includegraphics[scale=0.21]{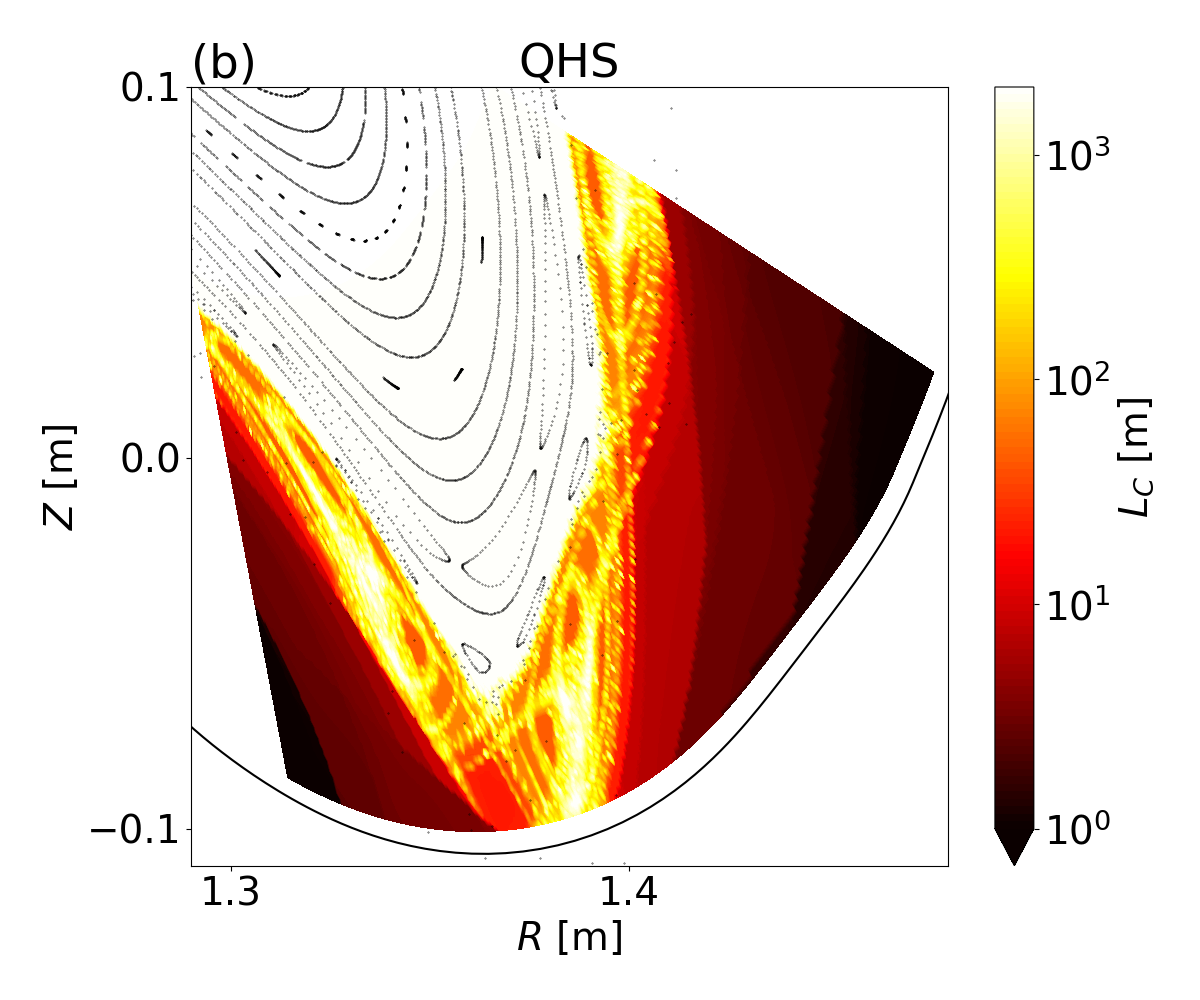}
    \includegraphics[scale=0.21]{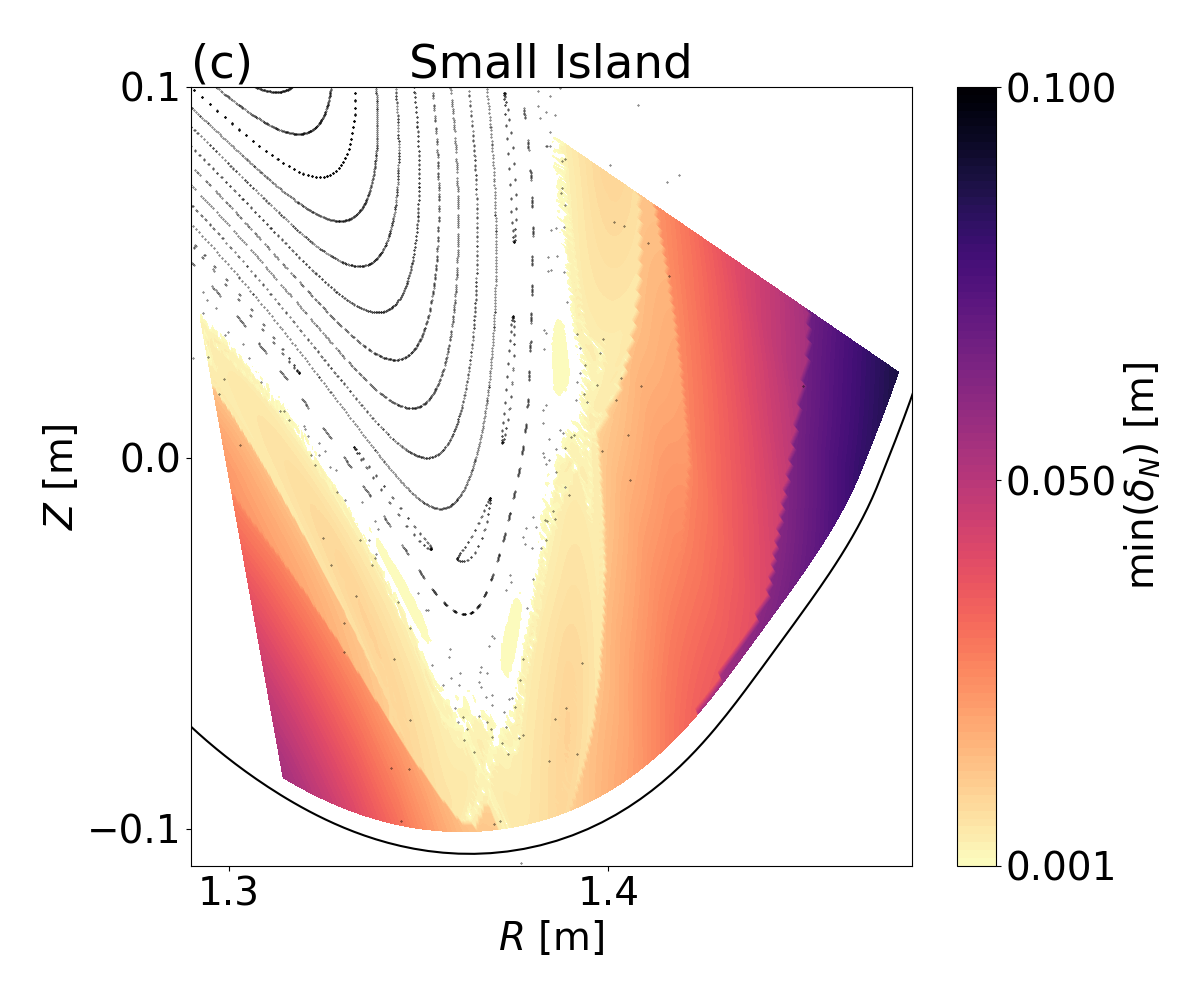}
    \includegraphics[scale=0.21]{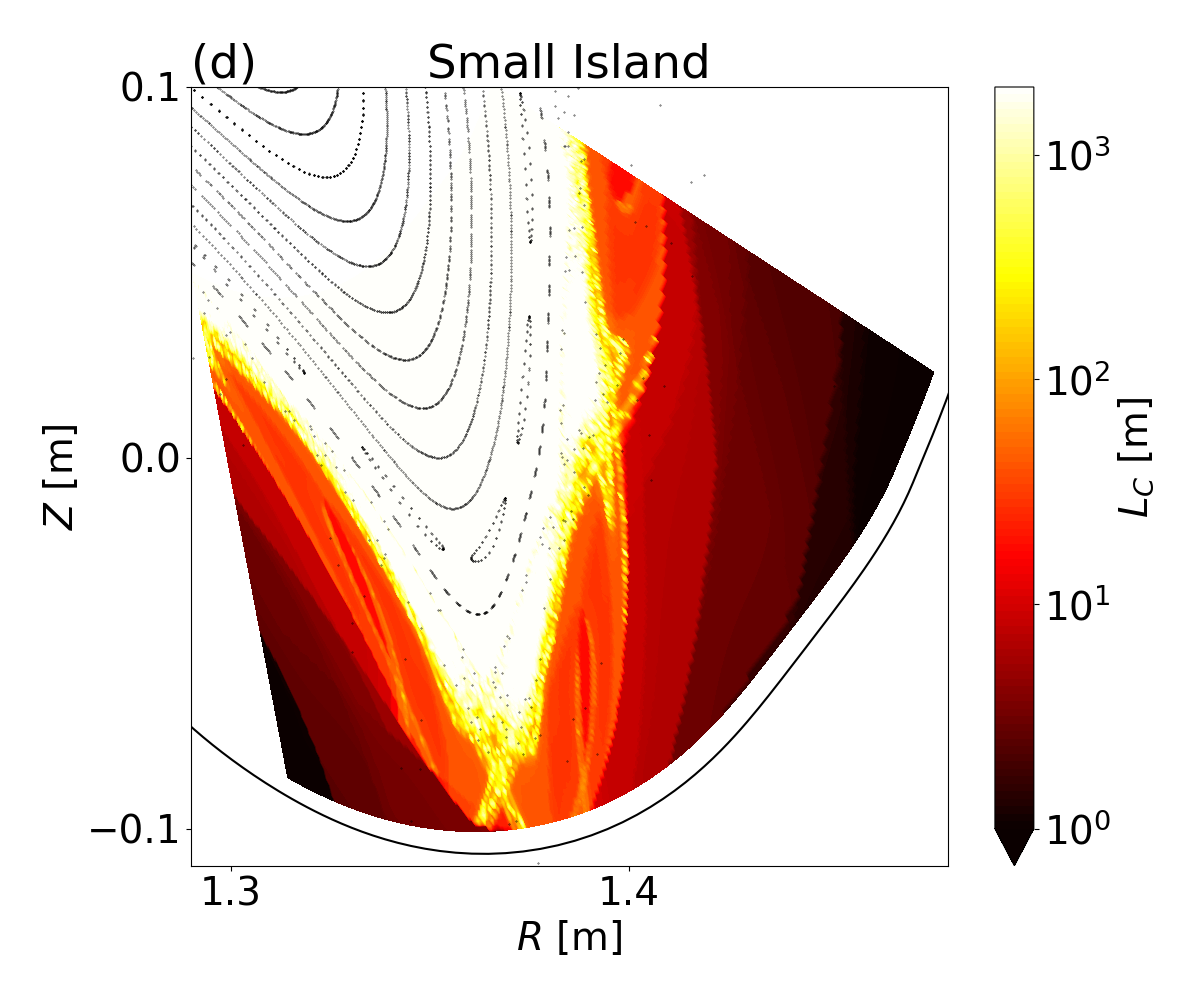}
    \includegraphics[scale=0.21]{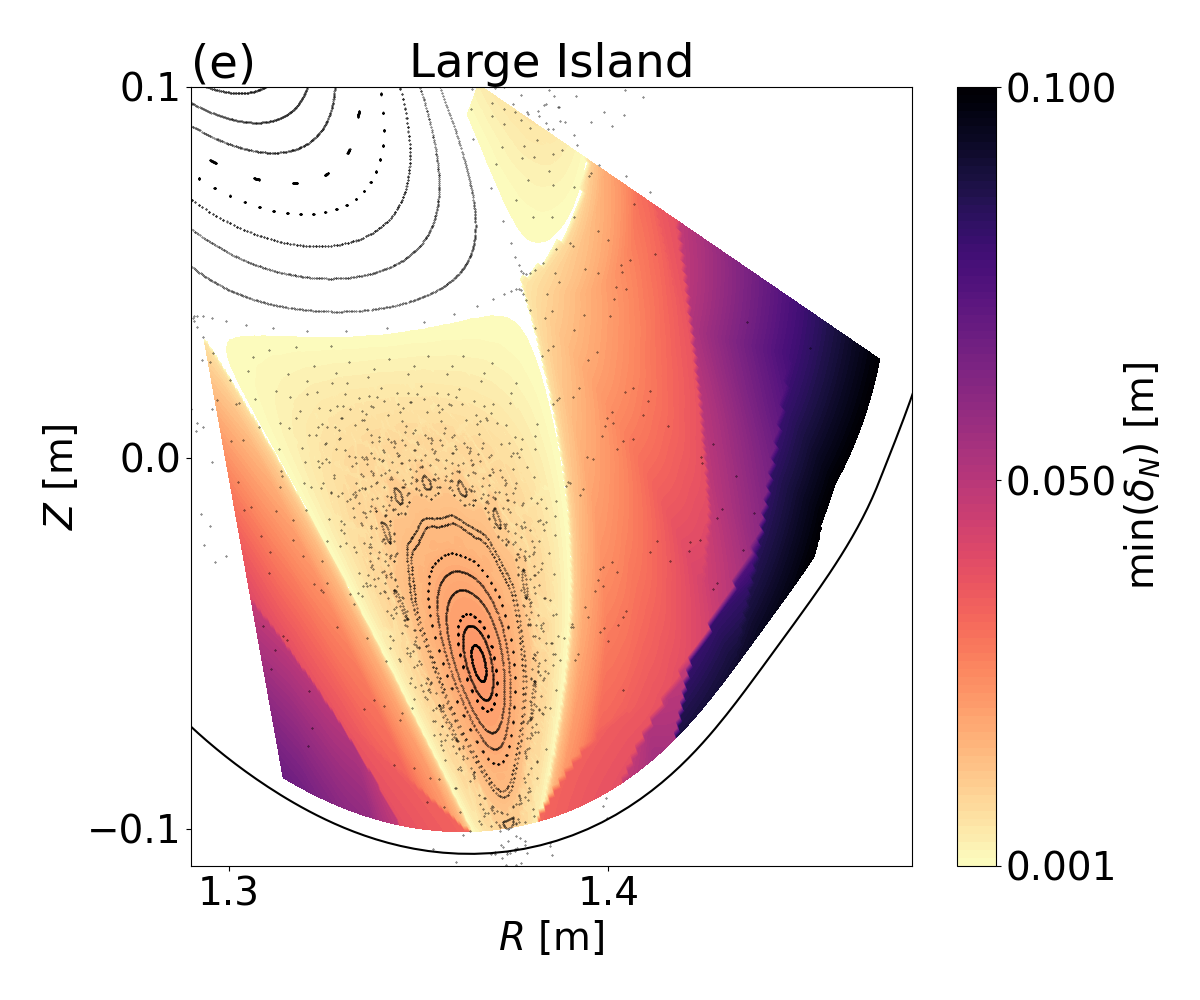} 
    \includegraphics[scale=0.21]{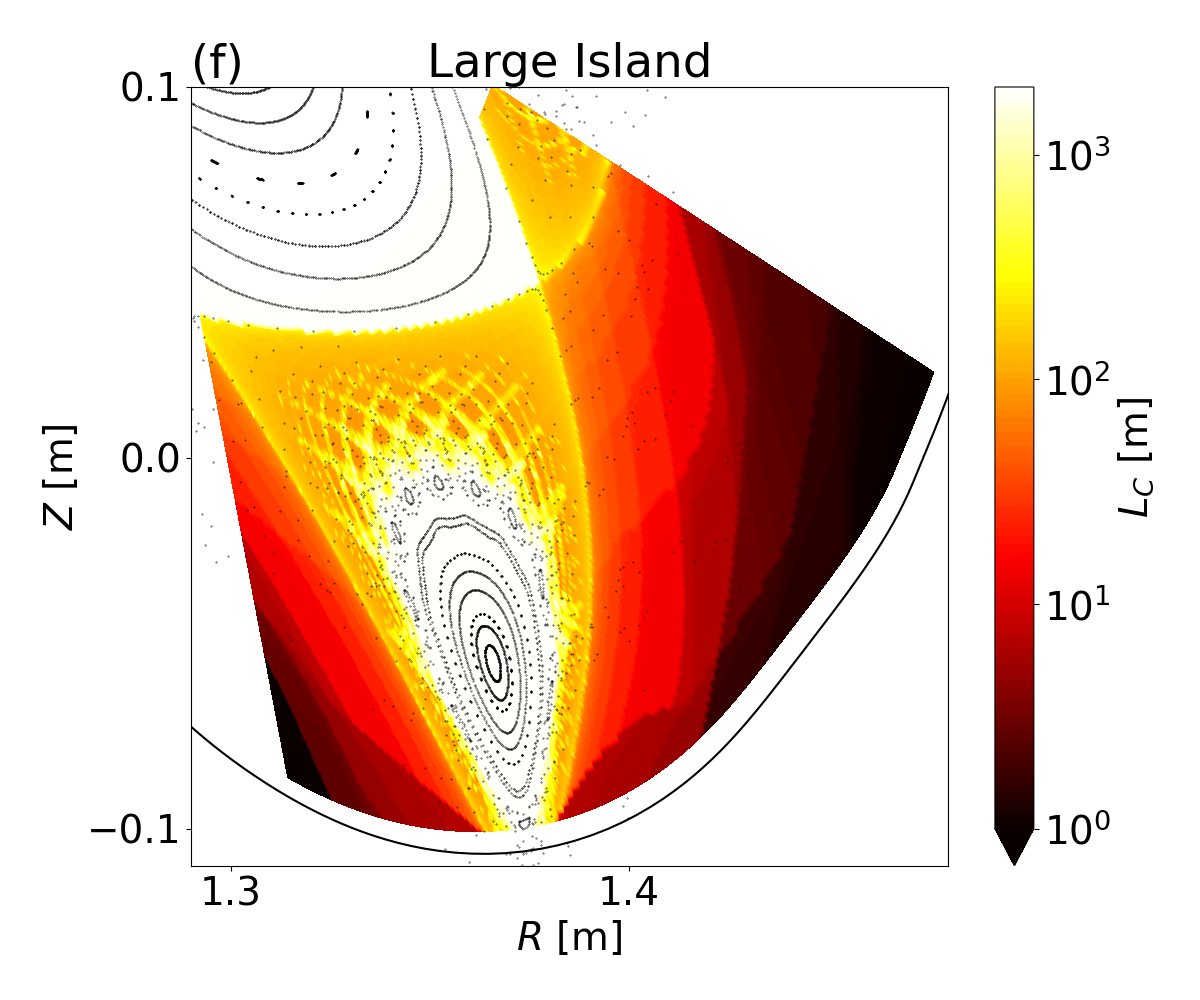}
    \includegraphics[scale=0.21]{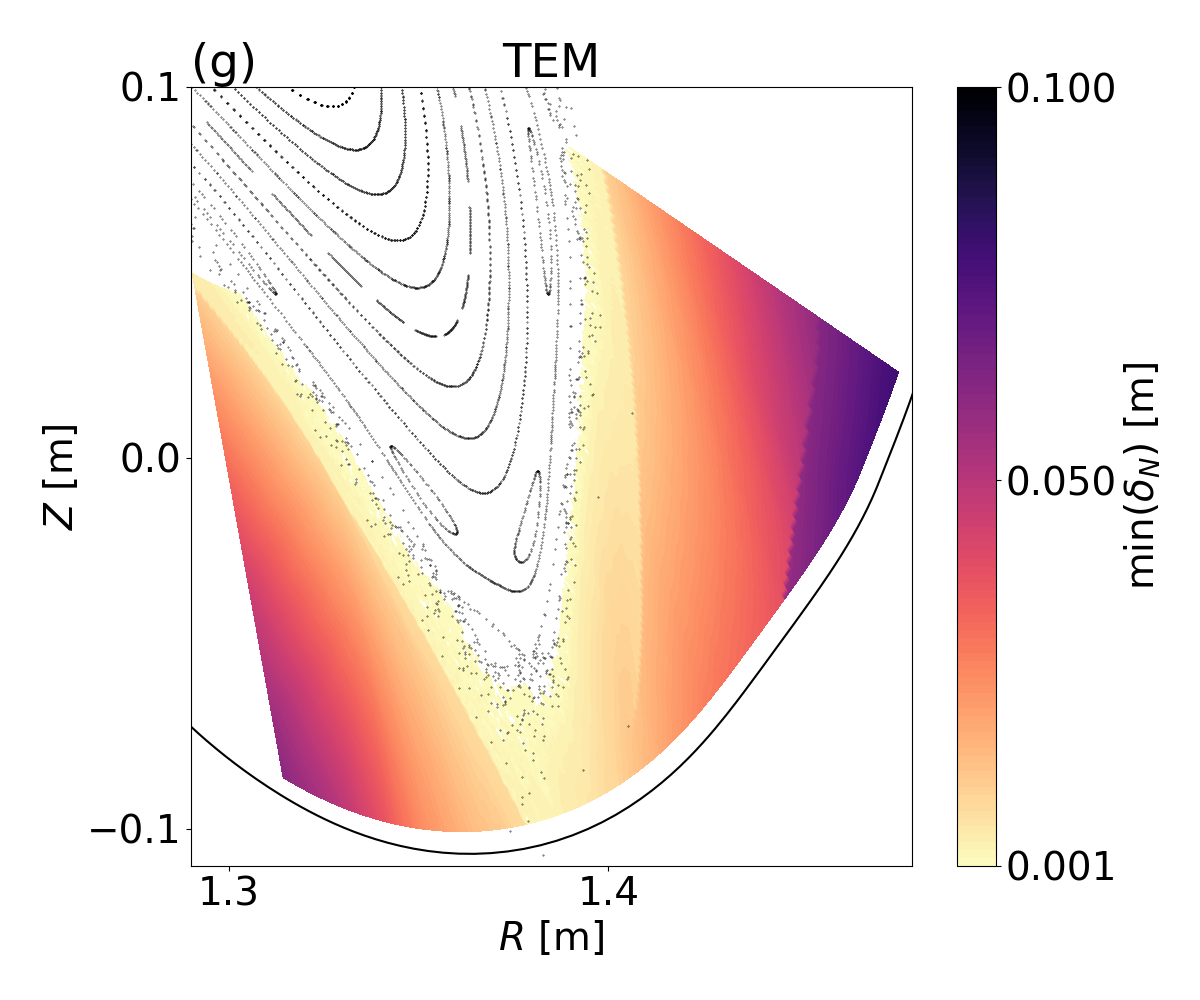}
    \includegraphics[scale=0.21]{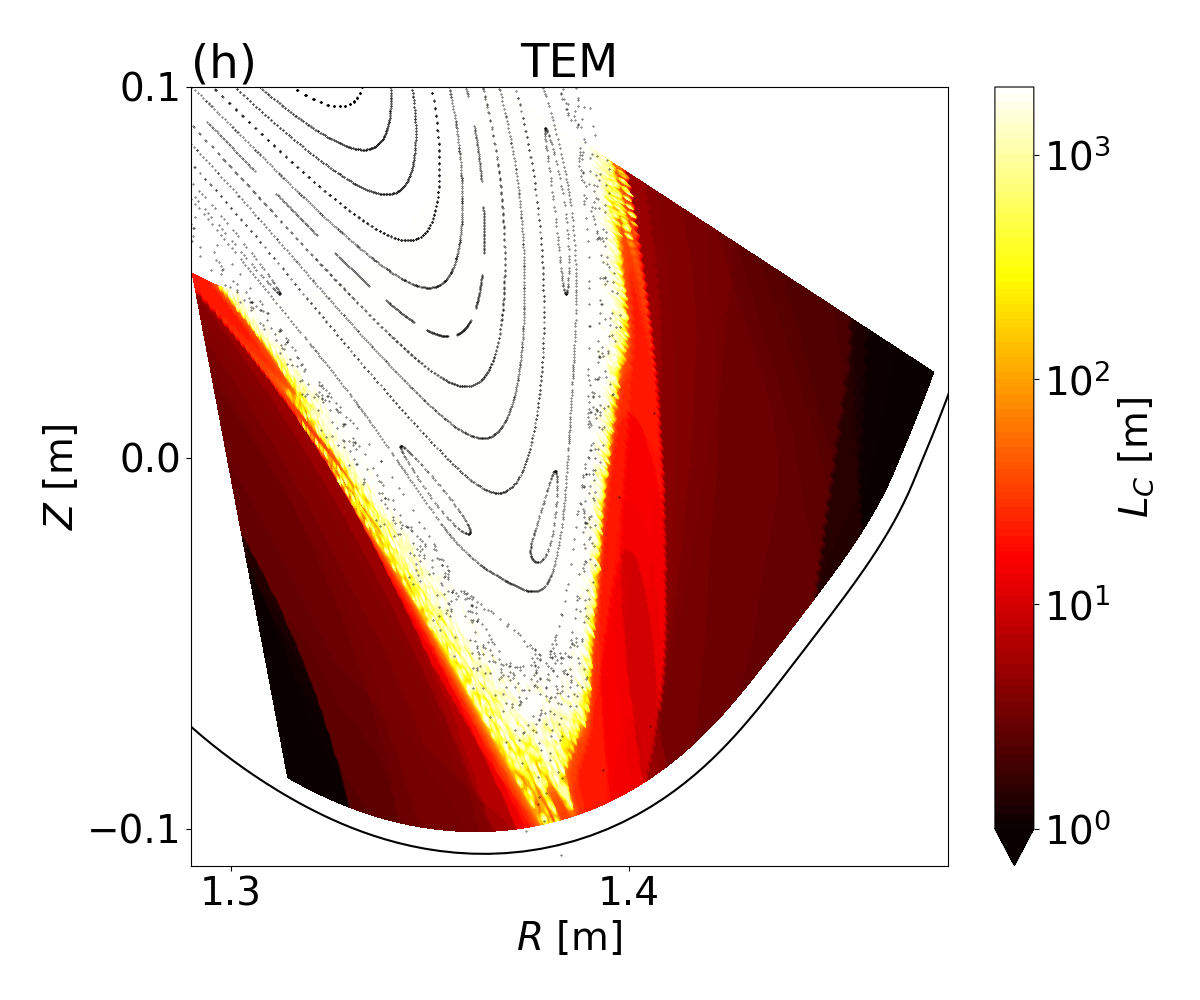}
    \caption{High resolution figures of the PWI corresponding to $\phi=18^\circ$ for each magnetic configuration. The left column is the radial connection and the right column is the connection length.}
    \label{fig:poincare18_Lc_DeltaN}
\end{figure}

The previous sections convey that the relationship between $L_C$ and min$(\delta_N)$ of the field lines near the wall is an inverse one. We aim to quantify this relationship further in this section and the results will associate the features of $L_C$ and min$(\delta_N)$ with resonant islands, cantori, and turnstiles. These results will guide the differences in the PWI across the different magnetic configurations which are emphasized through figures \ref{fig:high_res} - \ref{fig:poincare18_Lc_DeltaN}. 

The plots in figure \ref{fig:high_res} are a higher resolution calculation of figures \ref{fig:Lc_target} and \ref{fig:PsiN_target} focusing on the regions of $0^{\circ} \leq \phi \leq 25^{\circ}$ and $240^\circ \leq \theta \leq 315^\circ$. The left column is a contour map of min$(\delta_N)$ and the right column displays the corresponding $L_C$ for each case where the rows correspond to one configuration ((a) and (b) are QHS, for example). In these simulations, the computation was extended for a maximum $L_C = 10 \ $km. The two localized regions of high $L_C$ and low min$(\delta_N)$ (one between $0^\circ \leq \phi \leq 7^\circ$ and the other at $15^\circ \leq \phi \leq 22^\circ$) correspond with one another for each magnetic configuration. The details of these localized regions, however, show that there are subtle differences in each configuration's PWI. For example, between $0^\circ \leq \phi \leq 7^\circ$ the high $L_C$ region is comprised of helical striations in the QHS (figure \ref{fig:high_res} (a) and (b)), small island (figure \ref{fig:high_res} (c) and (d)), and TEM (figure \ref{fig:high_res} (g) and (h)) cases in contrast to the saturated $L_C$ appearance which has less sub-structure at very high $L_C$ in the large island case (figure \ref{fig:high_res} (e) and (f)). The high $L_C$ helical band region comprised of small-scale $L_C$ features exhibit less localized connection length in general. This is also seen in the toroidal range of $15^{\circ} \leq \phi \leq 22^{\circ}$ for the small island and TEM cases. It will be shown that the regions of very localized and high $L_C$ for both mentioned toroidal ranges of the large island case are due to the edge islands intercepting the lofted wall. This also occurs in the QHS case for $15^\circ \leq \phi \leq 22^\circ$ which has localized and high $L_C$. The described high $L_C$ features of the magnetic footprint are manifested similarly in the min$(\delta_N)$ calculation where regions of high $L_C$ correspond to low min$ (\delta_N)$. To emphasize the differences between each case further, we consider figure \ref{fig:2D_zoom_log_10^4_higherres}. 

Figure \ref{fig:2D_zoom_log_10^4_higherres} plots $L_C$ and min$(\delta_N)$ for a limited range of $\theta$ for only two toroidal angles from each plot of figure \ref{fig:high_res}. Each row of figure \ref{fig:2D_zoom_log_10^4_higherres} plots one quantity ((a) and (b) are $L_C$ and (c) and (d) are min$(\delta_N)$) while each column is one toroidal angle ((a) and (c) are $\phi=5^\circ$ and (b) and (d) are $\phi=18^\circ$). The distribution of $L_C$ and min$(\delta_N)$ are qualitatively characterized by a staircase-like distribution for $L_C \lesssim 10^2 \ $m. This is an indication of the different nested connection length layers present in the plasma edge and has been seen in similar plots of $L_C$ in devices such as Heliotron J \cite{cai_impact_2024}, W7-AS \cite{grigull_first_2001}, W7-X \cite{hammond_drift_2019}, and LHD \cite{morisaki_characteristics_2000}. Reference \cite{cai_impact_2024} shows the existence of a multi-fold layer consisting of constant $L_C$ layers embedded within the chaotic structure in the plasma edge. Hence, this behavior of $L_C$, which is also associated with min$(\delta_N)$, is consistent with what is seen in past work in different experimental devices.

Along with the aforementioned trend, figure \ref{fig:2D_zoom_log_10^4_higherres} also demonstrates the differences in the influence of topological structures on the PWI, particularly for the areas of high $L_C$ and low min$(\delta_N)$. Figure \ref{fig:2D_zoom_log_10^4_higherres} (a) and (c) focus on the poloidal region of $260^{\circ} \leq \theta \leq 280^{\circ}$ at $\phi = 5^{\circ}$ while figure \ref{fig:2D_zoom_log_10^4_higherres} (b) and (d) focus on the poloidal region of $280^{\circ} \leq \theta \leq 300^{\circ}$ at $\phi = 18^{\circ}$ from figure \ref{fig:high_res}. As this calculation was performed for a maximum $L_C = 10 \ $km, the only configuration with $L_C = 10 \ $km is the large island configuration (red) around $\theta \sim 290^{\circ}$ in figure \ref{fig:2D_zoom_log_10^4_higherres} (b) at $\phi = 18^\circ$. This is where one of the $4/4$ islands intersects the target surface. This is the only configuration with well-formed islands intersecting the lofted wall. The QHS configuration (orange) also has islands intersecting the wall, but they are not well-formed. Hence, the connection lengths are not as long in this region and are roughly up to $\mathcal{O} (10^2 \ $m) $- \mathcal{O} (10^3 \ $m). This is similarly observed in the small island configuration (green) where the edge islands are even less formed and the maximum connection length is $ \mathcal{O} (10^2 \ $m). Finally, the TEM case (blue) features long connection length field lines up to $\mathcal{O} (10^4 \ $m) in the edge plotted in figure \ref{fig:2D_zoom_log_10^4_higherres} (a). This configuration has a large plasma volume extending toward the lofted wall where the edge region has open field lines from overlapping resonant islands. This can explain the presence of very high $L_C$ close to the wall in the TEM case. The described topological features intersecting the lofted wall with different values of $L_C$ are also emphasized in figures \ref{fig:poincare5_Lc}, \ref{fig:poincare18_Lc}, \ref{fig:poincare5_Lc_DeltaN}, and \ref{fig:poincare18_Lc_DeltaN}.

Figures \ref{fig:2D_zoom_log_10^4_higherres} (c) and (d) of the min$(\delta_N)$ for all four configurations reveal that, although high $L_C$ is correlated with low min$(\delta_N)$, the connection is nuanced. For example, the large island configuration has very large $L_C$ at $\phi=5^\circ$ and $\theta \sim 265^\circ$ in figure \ref{fig:2D_zoom_log_10^4_higherres} (a) with min$(\delta_N) \sim 10^{-2} \ $m in figure \ref{fig:2D_zoom_log_10^4_higherres} (c). The two smallest min$ (\delta_N) < 10^{-2} \ $m in figure \ref{fig:2D_zoom_log_10^4_higherres} (c) occur outside the regions of very long connection length in figure \ref{fig:2D_zoom_log_10^4_higherres} (a) for $L_C \sim 10^2 \ $m. This is similarly observed in figure \ref{fig:2D_zoom_log_10^4_higherres} (d) where min$ (\delta_N) \sim 10^{-2} \ $m is wedged between two min$ (\delta_N)$ minimums. This behavior is not quite present in the other magnetic configurations. It will be shown in figure \ref{fig:Lc_vs_DeltaN} that this difference is due to island-dominated PWI in the large island case in contrast to the other cases. 

\begin{table}[H]
    \centering
    \begin{tabular}{|c | c | c |}
        \hline 
         figure & $a$ & $b$ \\
         \hline \hline 
         \ref{fig:Lc_vs_DeltaN} (a) & $-1.71 \pm 0.02$ & $0.006 \pm 0.0007$  \\
         \ref{fig:Lc_vs_DeltaN} (b) & $-1.74 \pm 0.03$ & $0.006 \pm 0.0008$ \\
         \hline 
    \end{tabular}
    \caption{Estimated values of $a$ and $b$ for figures \ref{fig:Lc_vs_DeltaN} (a) and (b) assuming the relationship: $L_C = b \ $min$(\delta_N)^a$. This fit is only performed for the small island configuration dataset.}
    \label{tab:powerlaw}
\end{table}

Figure \ref{fig:Lc_vs_DeltaN} interprets figure \ref{fig:2D_zoom_log_10^4_higherres} by plotting $L_C$ versus min$(\delta_N)$ for (a) $\phi=5^\circ$ and (b) $\phi=18^\circ$. The 4 configurations shown demonstrate the overall trend that $L_C$ and min$(\delta_N)$ have an inverse relationship. To describe this trend more quantitatively, a line fitting the small island configuration data (green) is plotted in black. This fit assumes a power law $L_C = b \ $min$(\delta_N)^{a}$ where $a$ and $b$ are parameters estimated in table \ref{tab:powerlaw} for each plot in figure \ref{fig:Lc_vs_DeltaN}. The small island configuration dataset was chosen for the fitting because this produced the highest quality power law. Additionally, power laws have been used to describe the behavior field lines in the edge for NRDs \cite{boozer_simulation_2018, punjabi_simulation_2020, punjabi_magnetic_2022}. The aim is to explore if a power law can distinguish varying PWI across the chosen magnetic configurations. It will be shown that the other cases have features which do not follow the chosen power law and will be described later in this section. 

The small island case features an X-point very close to the wall as seen in figures \ref{fig:poincare5_Lc_DeltaN} (c)-(d) and \ref{fig:poincare18_Lc_DeltaN} (c)-(d) and follows this power law behavior. The power law fitting is performed on this configuration's data and the calculated fitting parameters are tabulated in table \ref{tab:powerlaw}. The values show that the exponential parameter $a$ at the two different toroidal angles correspond with one another within a margin of error. The divertor legs of the X-point plotted in figures \ref{fig:poincare5_Lc_DeltaN} (c)-(d) and \ref{fig:poincare18_Lc_DeltaN} (c)-(d) clearly intercept the lofted vessel. This PWI dominated by these divertor legs is another confirmation of the field line min$ (\delta_N)$ behavior along the separatrix.  

The first observation in figure \ref{fig:Lc_vs_DeltaN} (a) and (b) is that the configuration with the largest deviation from a power law is the large island case. The $L_C$ versus min$(\delta_N)$ relationship demonstrates that there is an apparent bifurcation in the behavior where very high $L_C$ values have min$(\delta_N) \sim 10^{-2} \ $m and do not follow the trend of the fitted power law in both figure \ref{fig:Lc_vs_DeltaN} (a) and (b). This feature is somewhat weakly manifested in the QHS case in figure \ref{fig:Lc_vs_DeltaN} (b) for $\phi=18^\circ$. To explain this feature, figures \ref{fig:poincare5_Lc_DeltaN} and \ref{fig:poincare18_Lc_DeltaN} plot min$(\delta_N)$ and $L_C$ with an overlayed Poincar\`e map in black for $\phi=5^\circ$ and $\phi=18^\circ$, respectively, in the vicinity of the lofted wall. It is evident in the large island case in figures \ref{fig:poincare5_Lc_DeltaN} (e)-(f) and \ref{fig:poincare18_Lc_DeltaN} (e)-(f) that within an island $L_C \xrightarrow{} \infty$. Hence, a field line that is started within an island will always remain separated from the LCFS and therefore have longer min$(\delta_N)$. However, field lines about the X-point and along the divertor leg will have smaller min$(\delta_N)$ because these field lines are diverted and quickly hit the PFC. This explains why in figure \ref{fig:Lc_vs_DeltaN} the large island configuration has points in each branch of this bifurcated behavior depending on if the field line is near the leg of the X-point. This is similarly observed in the QHS case at $\phi=18^\circ$ of this figure, but because the edge island is not well formed, the field lines' $L_C$ are not as long as the field lines within the islands of the large island case. 

The TEM configuration has the largest plasma volume out of all the cases and meaning that its LCFS is closest to the wall. Therefore, this configuration should have the lowest average value of min$(\delta_N)$.  This is what is seen in figure \ref{fig:Lc_vs_DeltaN} where the blue points lie below those of the other configurations.

At first glance, the TEM configuration also appears to follow a power law. However, it also features a bifurcation seen in both figures \ref{fig:Lc_vs_DeltaN} (a) and (b) around min$ (\delta_N)$ of $10^{-2}$ m. The points from this dataset which follow the power law can be attributed to field lines near the LCFS which intercept the wall. Figures \ref{fig:poincare5_Lc_DeltaN} (g)-(h) and \ref{fig:poincare18_Lc_DeltaN} (g)-(h) also show that outside the LCFS, the field lines appear quite stochasticized which may be due to the presence and overlap of many high order edge islands. Much like the QHS and small island configuration, these islands are not well-formed. Therefore, these field lines are not expected to have very high $L_C$ and instead have larger min$ (\delta_N)$ due to the increased distance from the LCFS.

\section{Discussion and Conclusion}
\label{section:disc}

The studied magnetic configurations in this work result in qualitatively very similar magnetic footprints and radial connection patterns near the PFC. This confirms the previously calculated resilient edge field line behavior in HSX but now using an expanded vessel wall along with an additional metric, min$(\delta_N)$. Specifically, the regions of high $L_C$, which are the anticipated heat and particle flux regions, agree in general with low min$ (\delta_N)$ despite the vastly different topological features present across the configurations. Introducing the minimum distance to the LCFS min$ (\delta_N )$ of the edge field lines provides insight regarding the structures responsible for the PWI across the different magnetic configurations. Moreover, the relationship between $L_C$ and min$(\delta_N)$ reveal that the presence of large islands near the PFC impact the PWI differently than the other cases. 

The field lines with largest values of connection length confirm what is described in reference \cite{boozer_needed_2024} where field lines outside and nearest the LCFS which take many transits $L_C \xrightarrow{} \infty$ until they impact the PFC are interacting with cantori. However, field lines in the vicinity of an edge island also behave similarly with $L_C \xrightarrow{} \infty$. Figure \ref{fig:Lc_vs_DeltaN} shows that the difference between island and cantori field line behavior may be manifested in the minimum radial connection of the field line with respect to the LCFS. Namely, the field lines which do not follow the power law with min$ (\delta_N) \sim 10^{-2} \ $m as their $L_C$ becomes longer and longer are within an edge island. The field lines which follow a power law are close to the separatrix where very long $L_C$ combine with very small min$(\delta_N)$ may stay close to the LCFS or separatrix due to interaction with cantori. The study of field lines behavior crossing cantori and escaping through turnstiles has been studied or NRDs in references \cite{boozer_simulation_2018,punjabi_simulation_2020,punjabi_magnetic_2022} where the probability of field lines lost to the wall through turnstiles assumes a power law. Thus, employing this power law to describe the $L_C$ trend of the field lines in order to distinguish topological edge structures is relevant for studying field line behavior for NRDs in HSX. 

Despite these noted field line behavior differences, it is observed that the PWI is resilient across all cases. This is in contrast to previous work in \cite{bader_hsx_2017} where the large island case was observed to not exhibit resilient behavior. The difference here is that an expanded vessel wall was implemented as the PFC. The results demonstrate that if a large island is present and the wall or divertor target is outside of the island or separatrix, then resilient PWI is observed. This is similarly seen in reference \cite{strumberger_magnetic_1992}, and is different than what is observed in the W7-X island divertor configuration or in reference \cite{bader_hsx_2017} where the PFC in both cases deeply intersects the islands. Such effects of island divertor SOL geometry in W7-X has recently been shown to influence detachment behavior \cite{winters_first_2024} and lack of efficient neutral exhaust \cite{boeyaert_analysis_2023}. Thus, future work in resilient divertors in general must simulate divertor performance, \hb{such as what is done in \mbox{\cite{boeyaert_towards_2025}} investigating neutral particle exhaust in HSX with the lofted wall,} since NRD research to date has focused on the behavior of magnetic field lines in the plasma edge. This is of the utmost importance to assess the viability of the NRD for future stellarator fusion reactors. 

\ack 
This work was funded by the U.S. Department of Energy under grants DE-SC0023548, DE-SC0024548, DE-FG02-95ER54333, DE-SC0014210, and DE-FG02-93ER54222. This work was also supported by the Advanced Opportunity Fellowship (AOF) through the Graduate Engineering Research Scholars (GERS) and discretional funding by the College of Engineering at University of Wisconsin-Madison. Computing
time was provided through the University of Wisconsin - Madison’s Center for High Throughput Computing (CHTC).

\bibliographystyle{iopart-num}
% % Note the spaces between the initials

\section*{References}

\bibliography{paper}

\end{document}